\documentclass[final,onecolumn,12pt,journal,a4paper]{IEEEtran} 

\usepackage{epsfig,graphics}
\usepackage{subfigure}
\usepackage{amssymb,amsmath,amsfonts}
\usepackage[usenames]{color}
\usepackage{bbm}
\usepackage{psfrag} 
\usepackage{cite}
\usepackage{latexsym} 
\usepackage{amsmath}
\usepackage{multirow}
\usepackage{booktabs}

\IEEEoverridecommandlockouts 

\newcommand{\tz}[1]{{\color{black}#1}}

\title{Delay and Doppler Spreads of Non-Stationary Vehicular Channels for Safety Relevant Scenarios}

\author{\IEEEauthorblockN{Laura Bernad\'o, {\it Member, IEEE,} Thomas Zemen${^*}$, {\it Senior Member, IEEE,} Fredrik Tufvesson, {\it Senior Member, IEEE,} Andreas F.{} Molisch, {\it Fellow, IEEE,} Christoph F.{} Mecklenbr\"auker, {\it Senior Member, IEEE}}
\thanks{
This research was supported by the project NOWIRE funded by the Vienna Science and Technology Fund (WWTF) as well as the strategic FTW project I-0. The Austrian Competence Center FTW Forschungszentrum Telekommunikation Wien GmbH is funded within the program COMET - Competence Centers for Excellent Technologies by BMVIT, BMWFJ, and the City of Vienna. The COMET program is managed by the FFG. The research work described in this paper was carried out in cooperation within the COST Action IC1004 on Cooperative Radio Communications for Green Smart Environments. Part of this research has been presented in the \emph{IEEE 73rd Vehicular Technology Conference, VTC-Spring, 2011}.

L. Bernad\'o and T. Zemen ({\bf corresponding author}) are with Forschungszentrum Telekommunikation Wien (FTW), Vienna, Austria (e-mail: bernado@ftw.at, {\bf thomas.zemen@ftw.at}). F. Tufvesson is with the Department of Electrical and Information Technology, Lund University, Lund, Sweden (e-mail: Fredrik.Tufvesson@eit.lth.se). C. F. Mecklenbr\"auker is with the Institut of Telecommunications, Technische Universit\"at Wien, Vienna, Austria (e-mail: cfm@nt.tuwien.ac.at). A. F. Molisch is with the Department of Electrical Engineering, University of Southern California, Los Angeles, CA, USA (e-mail: molisch@usc.edu).
}}

\begin{document}

\maketitle

\begin{abstract}
Vehicular communication channels are characterized by a non-stationary time- and frequency-selective fading process due to rapid changes in the environment. The non-stationary fading process can be characterized by assuming \emph{local} stationarity for a region with finite extent in time and frequency. For this finite region the wide-sense stationarity and uncorrelated-scattering (WSSUS) assumption holds approximately and we are able to calculate a time and frequency dependent local scattering function (LSF). 
In this paper, we estimate the LSF from a large set of measurements collected in the DRIVEWAY'09 measurement campaign, which focuses on scenarios for intelligent transportation systems. We then obtain the time-frequency-varying power delay profile (PDP) and the time-frequency-varying Doppler power spectral density (DSD) from the LSF. Based on the PDP and the DSD, we analyze the time-frequency-varying root mean square (RMS) delay spread and the RMS Doppler spread. 
We show that the distribution of these channel parameters follows a bi-modal Gaussian mixture distribution. High RMS delay spread values are observed in situations with rich scattering, while high RMS Doppler spreads are obtained in drive-by scenarios. 
\end{abstract}

\section{Introduction}
Vehicular communication systems are currently under active investigation to enable intelligent transportation systems (ITS) contributing to safety improvements and environmentally friendly driving. The IEEE $802.11$p \cite{WAVEStandard} standard, now included in $802.11$ amendment $6$, has been drafted for that purpose. 

In order to evaluate the performance of vehicular communication systems, receiver structures are being developed and tested by means of numerical simulations \cite{Ivan2009, Lin2011, Kiokes2009, Nuckelt2011, Reichardt2011b, Bernado2010a, Zemen2012, Zemen12a, Fernandez12}. It is crucial to use realistic channel models for these simulations. Therefore, an effort in channel measurement and characterization has to be made in order to extract the channel parameters describing the fading properties of the radio channel. 

The main challenge for vehicular communications are the rapidly changing radio propagation conditions that strongly differ from cellular wireless networks. Both, the transmitter (TX) and the receiver (RX) are mobile, and the scattering environment can change rapidly. Hence, the vehicular communication channel is characterized by a non-stationary fading process \cite{Matz2005, Wilink2008, Bernado2008, Renaudin2010, Mecklenbraeuker2011, Molisch2009, Maurer2002, Acosta2004}.

\subsection*{Literature Review:}
In the literature vehicular channels are commonly characterized by means of the pathloss exponent \cite{Paier2007,Karedal2011,Paschalidis2011}, the RMS delay and Doppler spread \cite{Paier2007,Kunisch2008,Renaudin2008}, the distribution of the signal envelope, the power delay profile (PDP), and the Doppler power spectral density (DSD). Regarding the RMS delay spread, the smallest value is obtained in rural environments \cite{Kunisch2008,Tan2008}, and the largest in urban environments \cite{Tan2008,Renaudin2008,Sen2008}. Few results are available with respect to the RMS Doppler spread \cite{Maurer2002,Kunisch2008,Tan2008}. Convoy measurements show a lower RMS Doppler spread (but higher non power weighted spread) than on-coming measurements.
However, none of the mentioned publications analyze the time-variability of the statistical properties of the non-stationarity fading process in vehicular channels. 

The non-stationary fading process of vehicular channels can be characterized assuming local stationarity for a region with finite extent in time and frequency \cite{Matz2005, Matz2003a}. For this finite region the wide-sense stationarity and uncorrelated-scattering (WSSUS) assumption holds approximately and we are then able to calculate a time and frequency dependent local scattering function (LSF) \cite{Bernado2008, Paier2008}.

In \cite{Okonkwo2010} a related concept is used to calculate a LSF in order to quantify channel parameters and the channel capacity. However, the channel parameters are only characterized for a single stationarity region. The time-variability of the power spectral density is corroborated through numerical simulations obtained from a geometry based V2V channel model in \cite{Chelli2011}. However, the simulation results are not linked to, or validated by, empirical channel measurements.

The distribution of the stationarity distance of V2V measurement data is characterized in \cite{Renaudin2010}. Furthermore the distributions of large- and small-scale fading are extracted taking the stationarity distance into account. The empirical distributions are modeled by means of the generalized extreme value distribution \cite{Renaudin2010}.

Noteworthy is that the channel parameters reported in the previously listed publications were extracted from measurements using a cellular-based scenario description for highway, rural, suburban, and urban environments. These definitions are not well matched to the safety relevant scenarios where ITS will often be applied. A first preliminary evaluation of the RMS delay and RMS Doppler spread for only the \emph{in-tunnel} case was presented in \cite{Bernado2011b}.

\subsection*{Contributions of the Paper:}
{\begin{itemize}
\item We define vehicular radio propagation scenarios based on safety-critical applications for ITS. A large set of channel measurements is acquired in these scenarios.
\item The time-frequency-varying second central moment in the delay and in the Doppler domain are characterized by means of the RMS delay spread and RMS Doppler spread for all scenarios individually. 
\item We provide a simple but accurate model for the statistical distribution of these parameters using a bi-modal Gaussian mixture.
\end{itemize}}

\subsection*{Organization of the Paper:}
The measurement setup is introduced in Sec. \ref{se:MeasDescription}. In Sec. \ref{se:scenarios} we describe the measurement scenarios and their novel aspects. The measured data and the performed post-processing is discussed in Sec. \ref{se:analysis}. The mathematical description of the time-frequency-varying channel parameters is presented in Sec. \ref{se:TVparams}, and used for the full characterization of the channel parameters in Sec. \ref{se:statistics}. We draw conclusions in Sec. \ref{se:conclusions}.

\section{The DRIVEWAY'09 Measurement Campaign}
\label{se:MeasDescription}
This paper aims at statistically characterizing  the entire set of measurements collected in $2009$ in an extensive vehicle-to-vehicle radio measurement campaign, named DRIVEWAY'$09$.
The size and shape of the vehicle carrying the measurement equipment influences the measured channel frequency responses. Therefore we use passenger vehicles in order to obtain results representing real propagation conditions. The channel sounder and the batteries needed for power supply were loaded in the trunk of the TX and the RX vehicle.

The measurements were done with the RUSK-Lund channel sounder, which operates based on the switched sounding principle \cite{Thomae2000}. A multi-carrier sounding signal is utilized to obtain the time-variant channel estimates in the frequency domain \cite{RUSK}. The output of the channel sounder TX is connected, through a high speed switch, to a four-element uniform linear array mounted on the roof-top of a vehicle. The antennas are specially designed for this measurement campaign \cite{Klemp2010}. Similarly, the RX antennas on the roof of the second car are connected, through a switch, to the sounder RX. The array consists of four circular patch antennas separated by a distance of $\lambda /2$. Each antenna element is tuned in order to make its radiation pattern slightly directional, such that the main radiating directions of the antenna array are \emph{front, back, right}, and \emph{left} with respect to the driving direction. The transmit and receive antenna indices $n_\text{RX},n_\text{TX}\in\{1,2,3,4\}$ correspond to the main radiation directions $\{\text{left},\text{back}, \text{front}, \text{right}\}$. The combined antenna radiation pattern of the array resembles the one of an omnidirectional antenna.

The directionality of the patterns is achieved by setting the feeding point at an antenna edge, different for each element, and by the addition of a parasitic element, which acts as a director. The array and the parasitic elements are encapsulated into a conventional vehicular antenna module, with the particularity that it was mounted perpendicular to the driving direction in order to allow for directional analysis.

The antennas are matched to a center frequency of $5.6\,$GHz and a bandwidth of $240\,$MHz in accordance to the measurement frequency and bandwidth. The design of the antennas is such that the antenna gain remains on average within a variation of $10\,$dB for the whole bandwidth. More details regarding the antenna properties can be found in \cite{Klemp2010}.

\tz{In this paper we consider the antenna as well as the effects of the metallic vehicle hull as part of the channel. Only directionally resolved measurements would allow to separate the antenna effects, but this approach is not feasible for the measurement of time-variant non-stationary channels due to small coherence and stationarity times which limit the number of antenna elements.}

A total of $L= 4\times 4=16$ links are measured at a carrier frequency of $5.6$ GHz, which is the highest supported by the RUSK LUND channel sounder setup \cite{MEDAV}. Nevertheless, it is close enough to the $5.9$ GHz frequency band, assigned for ITS communications in Europe by the European commission \cite{ITS_2008_671_EC}. We expect that the propagation characteristics will not differ significantly at these two carrier frequencies. 

We measured a total bandwidth of $B=240$ MHz using a transmit power of $27$ dBm. The snapshot repetition time was set to $t_s=307.2\,\mu$s, with a sampling sequence length of $T=3.2\,\mu$s. Within one snapshot all $L=16$ antenna links are sounded individually by time multiplexing. After each sounding sequence of length $T$ a silent period of minimum length $T$ is applied to avoid distortions due to late reflections. This setup defines the maximum delay that can be resolved, hence $\tau_\text{max}=T$ \cite{RUSK}\footnote{\tz{The acquisition of a single snapshot requires $102.4\,\mu$s, then a waiting period was configured to obtain a snapshot repetition time of $t_s=307.2\,\mu$s. This approach was chosen such that with the given storage limitation of the channel sounder we were able to record measurement runs of $10$ seconds duration while on the other hand the temporal sampling is sufficient such that Doppler shifts up to $1.6$\,kHz can be resolved (caused by multiple reflections).}}. 

With these parameter settings we achieve a maximum resolvable Doppler shift of $1/(2\, t_s) = 1.6$ kHz, and a minimum delay resolution of $1/B=4.17$ ns. Table \ref{tab:01:01} summarizes the DRIVEWAY'09 main sounding parameters. More details regarding the measurement campaign can be found in \cite{Paier2010t}.

\begin{table}[htb!]
\begin{center}
\caption{DRIVEWAY'09 measurement campaign parameter details}
\label{tab:01:01}
\begin{tabular}{@{}ll@{}} \toprule
parameter             &   value  \\
\midrule
channel             & $4\times4$ MIMO \\

carrier frequency         & $5.6$\,GHz    \\

measurement bandwidth       & $240$\,MHz\\

transmit power          & $27$\,dBm     \\

sampling sequence length & $3.2\,\mu$s\\

snapshot repetition time     & $307.2$\,$\mu$s \\

recording time          & $10 \ldots 20$\,s \\
\bottomrule
\end{tabular}
\end{center}
\end{table}

\section{Safety-related Measurement Scenarios}
\label{se:scenarios}
\begin{figure}
\begin{center}
\includegraphics[width=0.85\columnwidth]{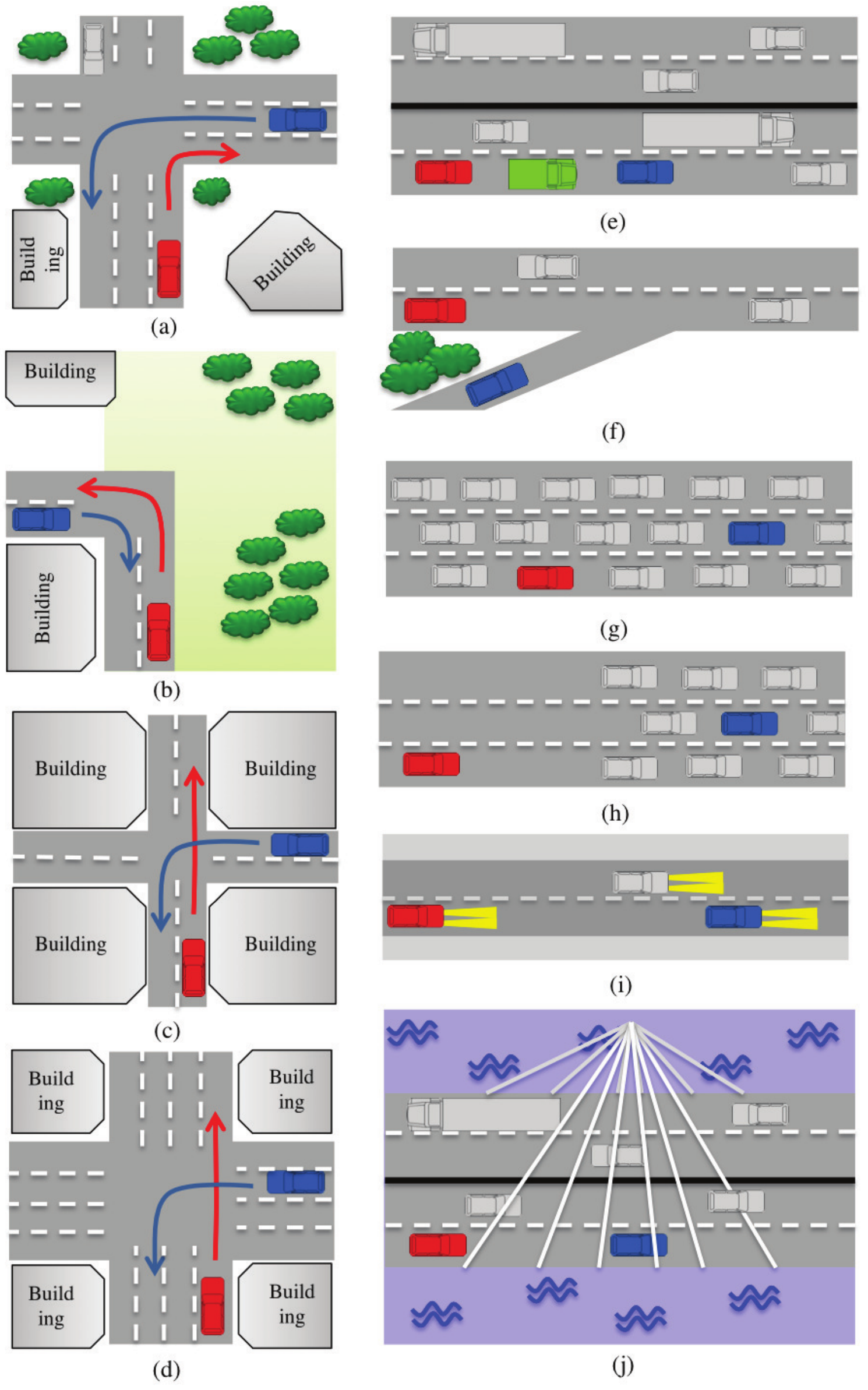}
\caption{DRIVEWAY'09 scenarios: (a) \emph{street crossing - suburban with traffic}, (b) \emph{street crossing - suburban without traffic}, (c) \emph{street crossing - urban single lane}, (d) \emph{street crossing - urban multiple lane}, (e) \emph{general LOS obstruction - highway}, (f) \emph{Merging lanes - rural}, (g) \emph{Traffic congestion - slow traffic}, (h) \emph{Traffic congestion - approaching traffic jam}, (i) \emph{In-tunnel}, (j) \emph{On-bridge}.}
\label{fig:scenarios}
\end{center}
\end{figure}

The measurements presented in this publication are defined for application specific scenarios, based on the European Telecommunications Standards Institute (ETSI) basic ITS application set definition \cite{ETSITR102638}. 
In the following we present these scenarios together with their particularities: 
\begin{itemize}
\item \emph{Road crossing}: This scenario consists of a conventional street crossing in rural, suburban, and urban environments, where both vehicles approach the crossing from perpendicular directions, driving at speeds between $2.8$ and $13.9\,$m/s ($10-50\,$km/h). Within this category we define four sub-scenarios so that we can measure crossings under various configurations: \emph{suburban with traffic} (Fig. \ref{fig:scenarios} (a)), \emph{suburban without traffic} (Fig. \ref{fig:scenarios} (b)), \emph{urban single lane} (Fig. \ref{fig:scenarios} (c)), and \emph{urban multiple lane} (Fig. \ref{fig:scenarios} (d)). The placement of buildings and therefore the availability of a LOS component will be an important aspect for the modeling of this scenario as we will see in Section~\ref{se:statistics}.

\item \emph{General LOS obstruction - highway} (Fig. \ref{fig:scenarios} (e)): This scenario investigates the influence of line-of-sight (LOS) obstruction in \emph{highway} environments. Both, the TX and RX are driving in the same direction on the highway with similar velocities between $19.4$ and $30.5\,$m/s ($70-110\,$km/h). There are other big trucks which are intermittently obstructing the LOS between TX and RX. 

\item \emph{Merging lanes - rural} (Fig. \ref{fig:scenarios} (f)): Here we consider a special intersection case; a ramp intersecting with a main street with partly obstructed junction. {V2V} communication occurs between vehicles driving in the same direction at $22.2-25\,$m/s ($80-90\,$km/h). 

\item \emph{Traffic congestion}: Different situations in a traffic congestion are considered here: \emph{Slow traffic}, where both, the TX and RX vehicle are stuck in a traffic jam driving, (Fig. \ref{fig:scenarios} (g)). They move in the same direction with slow velocities between $4.2$ and $8.3\,$m/s ($15-30\,$km/h), and are surrounded by other vehicles moving with similar speed. We also consider the case of \emph{approaching a traffic jam}, where the RX vehicle is stuck in a traffic jam (velocities close to $0\,$km/h) while the TX vehicle approaches from behind at around $16.7-19.4\,$m/s ($60-70\,$km/h), (Fig. \ref{fig:scenarios} (h)). 

\item \emph{In-tunnel} (Fig. \ref{fig:scenarios} (i)): Two vehicles drive in the same direction at more or less similar velocities between $22.2$ and $30.5\,$m/s ($80-110\,$km/h) inside a tunnel. The tunnel allocates only one driving direction with two lanes.

\item \emph{On-bridge} (Fig. \ref{fig:scenarios} (j)): Two vehicles drive in the same direction at around $27.8\,$m/s ($100\,$km/h) over a bridge with a separation of about 150 m. The bridge is composed of big metallic structures, equidistantly spaced. The specific bridge where measurements were performed is spanning the sea over more than $5\,$km.
\end{itemize}

We have carried out between 3 and 15 measurement runs per scenario in order to enable statistical analysis.

\section{Measured Data and Analysis}
\label{se:analysis}

In this section we present the mathematical notation for the measured impulse response and we describe the calculation of the time-frequency-varying power spectral density of the non-stationary fading process. Due to lack of space, we are going to augment the description of each parameter with only two illustrative examples taken from the DRIVEWAY'09 measurement campaign. These examples are the \emph{street crossing} of single lane streets in an \emph{urban} environment (Fig. \ref{fig:scenarios} (c)), and the \emph{general LOS obstruction} in a \emph{highway} environment (Fig. \ref{fig:scenarios} (e)). Other examples can be found in \cite{Bernado2012thesis}.

\subsection{Recorded Impulse Responses}
In vehicular communications, the environment changes rapidly due to the high mobility of the RX, the TX, and scatterers; and the antennas are placed on the roof-top of the vehicles. \tz{Under these conditions the channel frequency response $H(t,f)$ is time-varying and frequency-selective, where continuous time and frequency is denoted by $t$ and $f$, respectively.}

For the DRIVEWAY'09 measurements, $Q=769$ frequency bins in $B=240$ MHz total measured bandwidth are collected over $L=16$ links. For each measurement run, we record a total number of $S=32500$ or $S=65000$ snapshots with a snapshot repetition time of $t_\mathrm{s}=307.2\,\mu\text{s}$. \tz{Due to the multi-carrier (frequency-domain) principle of the channel sounder it directly supplies a sampled measurement of the time-variant frequency response:
\begin{equation}
H_\ell[m,q]\triangleq H_\ell(t_\text{s} m, f_\text{s} q)
\end{equation}
where the time index $m \in{\{0, \ldots,  S-1\}}$, the frequency index $q \in{\{0, \ldots,  Q-1\}}$, and link index $\ell \in {\{1, \ldots,  L\}}$. The resolution in the frequency domain is denoted by $f_\mathrm{s}=B/Q$.} The mapping between the link index $\ell$ and the TX and RX antenna indices is given by 
\begin{equation}
\ell= 4 (n_\text{TX}-1) + (5-n_\text{RX})\,,
\end{equation}
see also \cite[Fig. 5.5 and Tab. 5.2]{Bernado2012thesis}.

\subsection{Locally Defined Power Spectral Density}
In this paper we are especially interested to characterize the dispersion of the signal through the vehicular channel in the delay-Doppler domain caused by multipath propagation and mobility. Due to the fast changing propagation conditions, which are a key feature of vehicular communications, the observed fading process is non-stationary \cite{Matz2005,Wilink2008,Bernado2008,Renaudin2010}. Since the environment changes with a finite rate we can overcome the non-stationarity by approximating the fading process to be locally stationary for a region with finite extent in time and frequency \cite{Matz2003a}. This allows us to estimate locally the power spectral density of the fading process to describe its time-frequency-varying statistical behaviour and its Doppler-delay dispersion, also discussed in \cite{Bernado2012book}.\\ 
 
\begin{figure*}
\begin{center}
\subfigure[PDP - \emph{Road crossing - urban} intersection]{\includegraphics[width=0.5\textwidth]{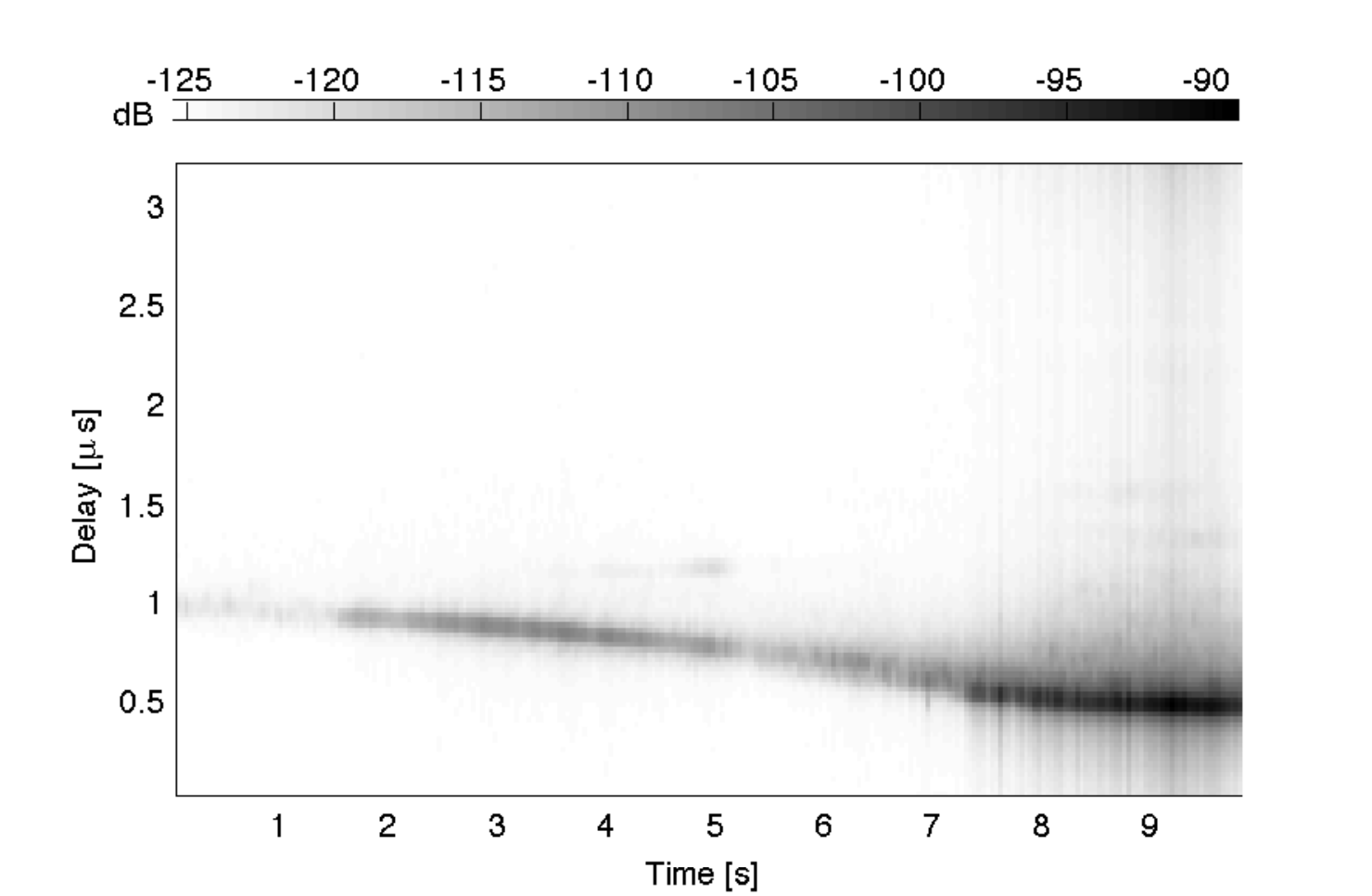}}%
\hspace{-0.5cm}
\subfigure[PDP - \emph{General LOS obstruction - highway} convoy]{\includegraphics[width=0.5\textwidth]{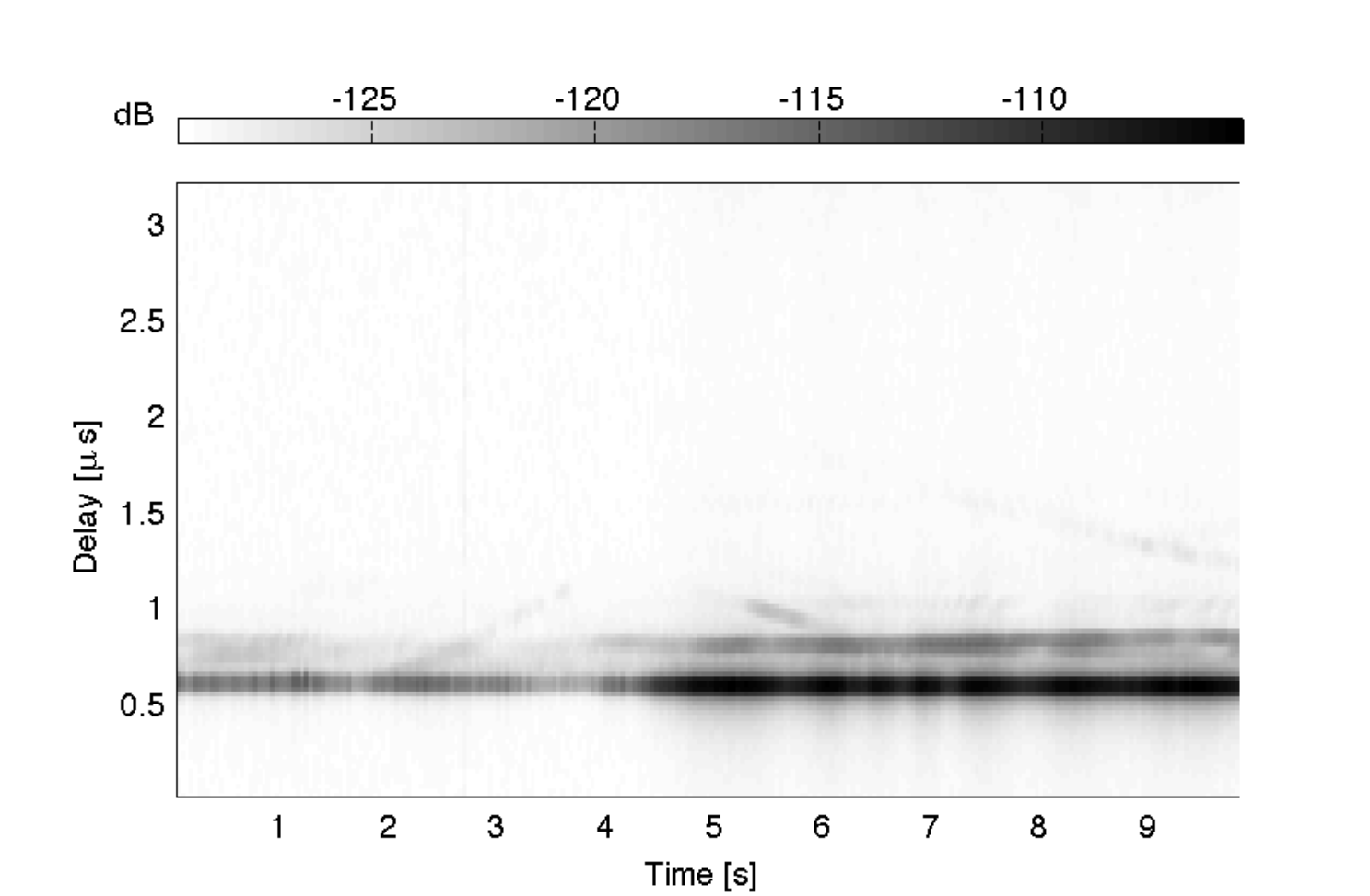}}%
\\
\subfigure[DSD - \emph{Road crossing - urban} intersection]{\includegraphics[width=0.5\textwidth]{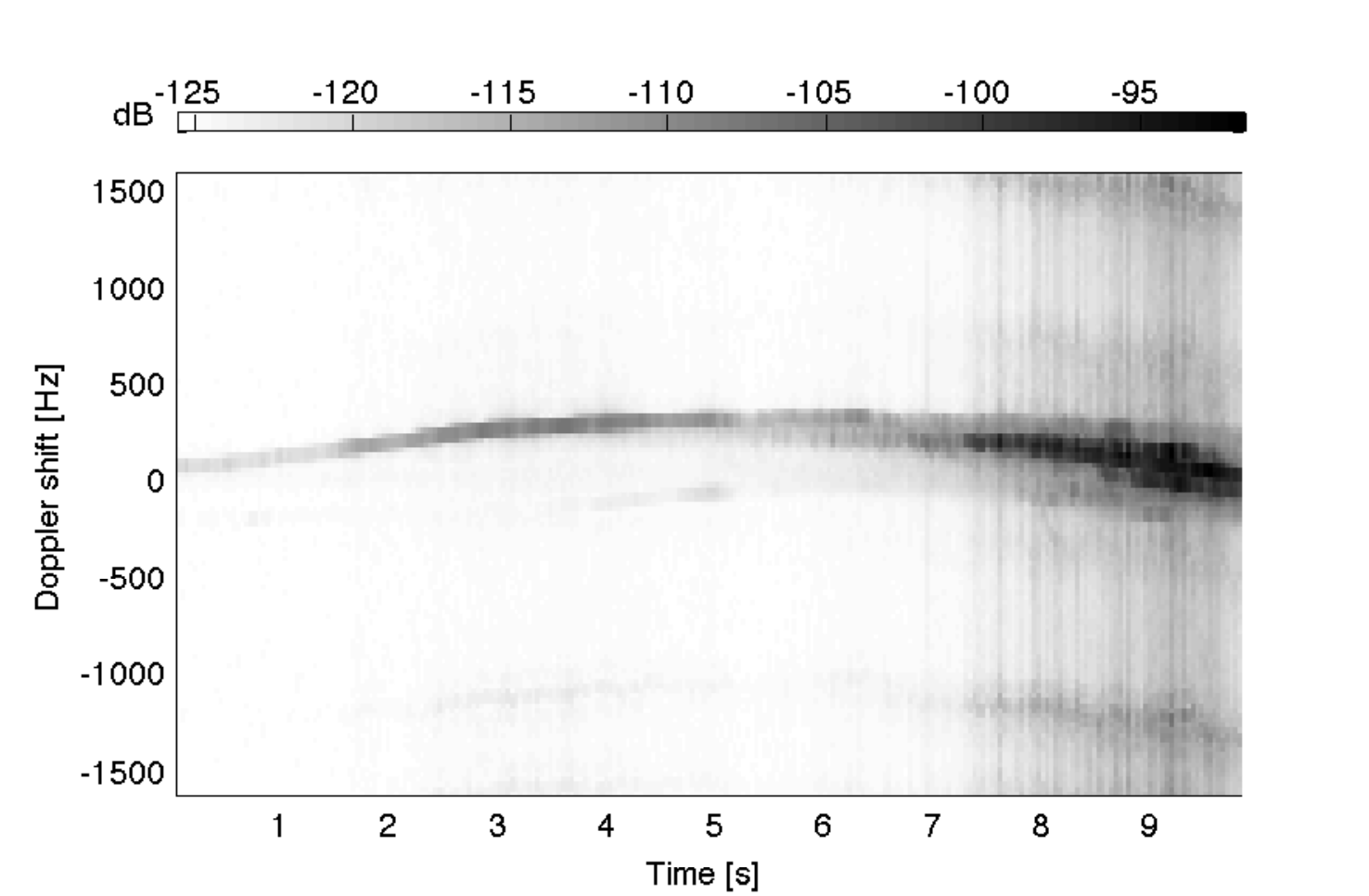}}
\hspace{-0.5cm}
\subfigure[DSD - \emph{General LOS obstruction - highway} convoy]{\includegraphics[width=0.5\textwidth]{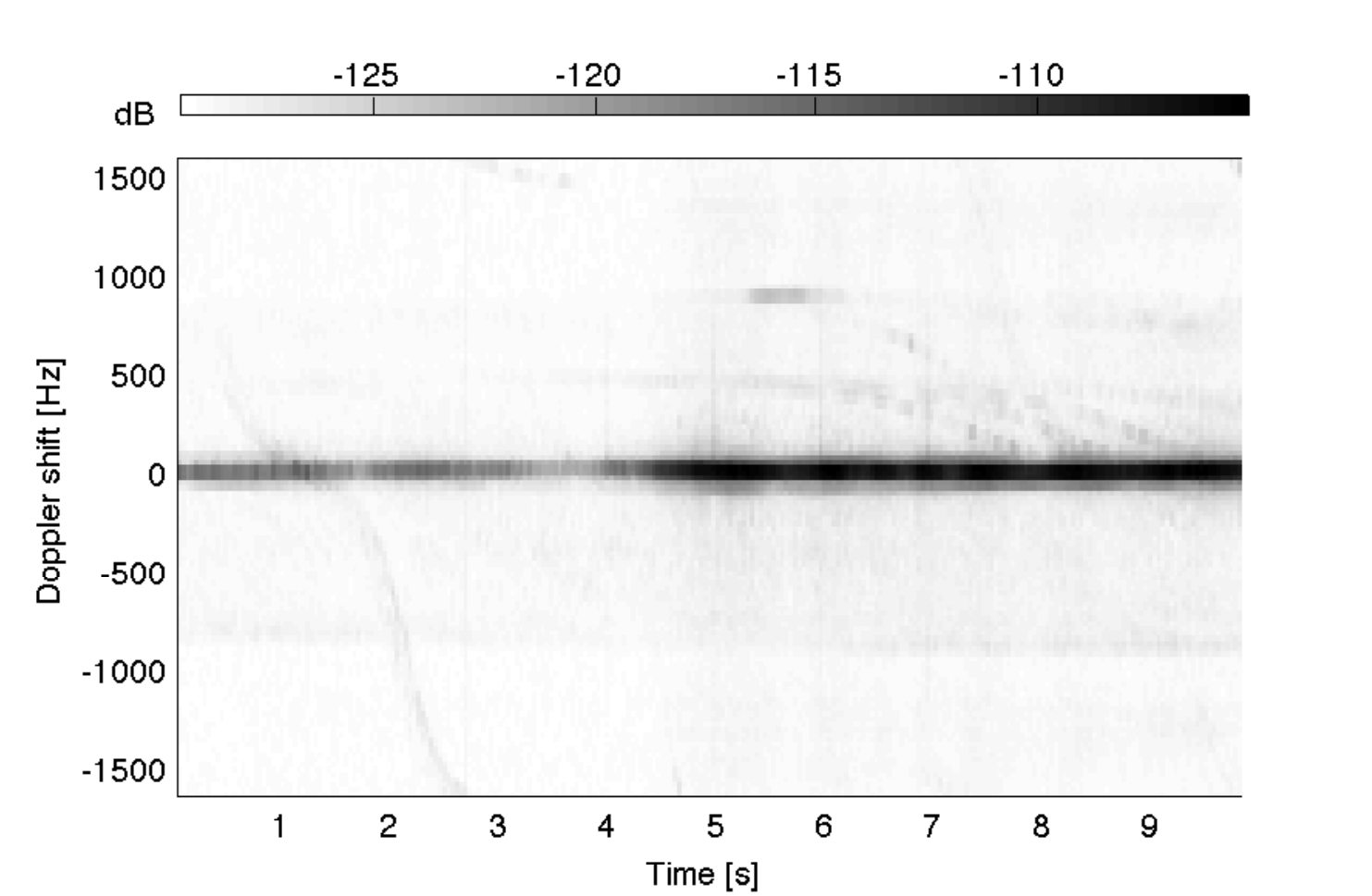}}
\caption{Time-varying power spectral densities for the frequency range $5480$-$5520\,$MHz from two different measurement scenarios: (a) and (c) \emph{{street crossing of two single lane streets in \emph{urban} environment}} measurement, velocities of approximately $2.8\,$m/s ($10\,$km/h); (b) and (d) \emph{general LOS obstruction in the highway}, convoy measurement with temporally obstructed LOS, constant velocities of $33.3\,$m/s ($120$\,km/h).}%
\label{fig:04:06}%
\end{center}
\end{figure*}
 
\subsubsection{The Local Scattering Function (LSF) Estimator}
\tz{When estimating the power spectrum of a process using measurements, it is very difficult to obtain statistically independent realizations of the same process. By tapering the measurement data using orthogonal windows, and by estimating the spectrum of each individual resulting windowed data set, we obtain multiple independent spectral estimates from the same measurement data set. The total estimated power spectrum is then calculated by averaging over all tapered spectra \cite{Percival1993,Thomson1982}. This approach allows to balance the noise variance and the square bias.}

For sampled time- and frequency selective channels $H[m,q]$ we use the discrete version of the LSF multitaper based estimator, introduced in \cite{Matz2003a, Matz2005} for the continuous-time case. The tapers in our application are computed from $I$ orthogonal time-domain tapers and $J$ orthogonal frequency-domain tapers, resulting in a total of $IJ$ orthogonal two-dimensional tapering functions.

We assume that the fading process is locally stationary within a stationarity region that has a general extent of $M \times N$ samples in time and frequency, respectively (later we will specify the concrete values for the extend of the stationarity region). Then, we calculate the LSF for consecutive stationarity regions. The time index of each stationarity region $k_t\in \{1 ,\ldots, \lfloor S/M-1\rfloor \}$ and corresponds to its center, the frequency index $k_f\in \{1 ,\ldots, \lfloor Q/N-1\rfloor \}$ is also corresponding to its center.
The relative time index within each stationarity region is $m'\in\{-M/2,\ldots,  M/2-1\}$. The relationship between the relative and absolute time index is given by $m=k_t \cdot M + m'$. Similarly, the relative frequency index within each stationarity region is $q'\in\{-N/2,\ldots,  N/2-1\}$, and its relationship to the absolute frequency index is given by $q=k_f \cdot N + q'$.

We compute an estimate of the discrete LSF \cite{Matz2005}, \cite{Paier2008} as
\begin{equation}
\label{eq:07:04}
\hat{\mathcal{C}}[k_t, k_f; n, p] = \frac{1}{IJ}\sum_{w=0}^{IJ-1}\left| \mathcal{H}^{(\Vec{G}_w)}[k_t, k_f; n, p] \right|^2
\end{equation}
where $n\in\{0,\ldots,N-1\}$ denotes the delay index, and $p\in\{-M/2,\ldots M/2-1\}$ the Doppler index, respectively. The LSF at ($k_t$,$k_f$) corresponds to the center value of the time-frequency stationarity region. The windowed frequency response reads
\begin{align}
&\mathcal{H}^{({G}_w)}[k_t, k_f; n, p] = \sum_{m'=-M/2}^{M/2-1} \sum_{q'=-N/2}^{N/2-1} \\
\nonumber & H[m'-k_t,q'-k_f] {G}_w[m',q'] \mathrm{e}^{-\mathrm{j} 2\pi (p m'-n q')},
\end{align}
The window functions ${G}_w[m',q']$ shall be well localized within the support region $[-M/2,M/2-1]\times[-N/2,N/2-1]$.  

We apply the discrete time equivalent of the separable window function used in \cite{Matz2003a}, $G_w[m',q'] = u_i[m'+M/2] \tilde{u}_j[q'+N/2]$ where $w=iJ+j$, $i\in\{0,\ldots,I-1\}$, and $j\in\{0,\ldots,J-1\}$. The sequences $u_i[m']$ are chosen as the discrete prolate spheroidal sequences (DPSS) \cite{Slepian1978} with concentration in the interval $\mathcal{I}_M=\{0,\ldots,M-1\}$ and bandlimited to $[-I/M, I/M]$.
The {DPSS} are the solutions to the Toeplitz matrix eigenvalue equation \cite{Slepian1978,Thomson1982}
\begin{equation}
\sum_{\ell=0}^{M-1}\frac{\sin(2\pi \frac{I}{M}(\ell-m'))}{\pi(\ell-m')} u_i[\ell]= \lambda_i u_i[m']\,.
\end{equation}
The sequences $\tilde{u}_j[q']$ are defined similarly with concentration in the interval $\mathcal{I}_N=\{0,\ldots,N-1\}$ and bandlimited to $[-J/N, J/N ]$ as
\begin{equation}
\sum_{\ell=0}^{N-1}\frac{\sin(2\pi \frac{J}{N}(\ell-q'))}{\pi(\ell-q')} \tilde{u}_j[\ell]= \lambda_j \tilde{u}_j[q']\,.
\end{equation}

In \cite{Bernado2012} we showed that the fading process in the vehicular radio channel is neither stationary in time (i.e. the wide-sense stationary (WSS) assumption does not hold) nor in frequency (i.e. the uncorrelated-scattering (US) assumption does not hold), and obtained the minimum stationarity region dimensions of $40\,$ms in time and $40\,\,$MHz in frequency. Therefore, for this investigation we select $M=128$, and $N=128$, which correspond to a stationarity region of $39.32\,$ms and $39.95\,$MHz. 

With these parameters the obtained resolutions are $t_\mathrm{s}=307.2\,\mu\text{s}$ in time, $f_\mathrm{s}=312.09$ kHz in frequency, $\tau_\mathrm{s}= Q/(BN) = 25\,$ns in delay and $\nu_\mathrm{s}=1/(t_sM)=25.43\,$Hz in the Doppler domain {respectively}.

\subsubsection{Power Delay Profile and Doppler Power Spectral Density}
\label{sec:PDPDSD}
The power delay profile (PDP) and the Doppler power spectral density (DSD) are widely used to describe the average dispersion of the transmitted signal in time and frequency. The PDP is the projection of the LSF on the delay domain, whereas the DSD is the projection of the LSF on the Doppler domain \cite{Matz2005, Hlawatsch11}.

Based on the LSF $\hat{\mathcal{C}}[k_t, k_f; n, p]$, the time-frequency-varying PDP and time-frequency-varying DSD can be defined as 
\begin{equation}
\hat{P}_{\tau}[k_t,k_f;n]={E}_{p}\{\hat{\mathcal{C}}[k_t,k_f; n, p]\}=\frac{1}{M}\sum_{p=-M/2}^{M/2-1}{\hat{\mathcal{C}}[k_t,k_f; n, p] }
\label{eq:PDP}
\end{equation}
and
\begin{equation}
\hat{P}_{\nu}[k_t,k_f;p]={E}_{n}\{\hat{\mathcal{C}}[k_t,k_f; n, p]\}=\frac{1}{N}\sum_{n=0}^{N-1}{\hat{\mathcal{C}}[k_t,k_f; n, p] },
\label{eq:DSD}
\end{equation}
where $E_x\{\cdot\}$ denotes expectation over the variable $x$.

We consider the combined LSF for $L=16$ links in order to resemble an omnidirectional antenna radiation pattern as 
\begin{equation}
\mathcal{\hat{C}}[k_t,k_f;n,p]=\frac{1}{L}\sum_{\ell=1}^{L}\mathcal{\hat{C}}^{(\ell)}[k_t,k_f; n, p],
\end{equation} 
where $\mathcal{\hat{C}}^{(\ell)}[k_t,k_f;n,p]$ is the LSF estimated for each individual link $\ell$ as in Eq. (\ref{eq:07:04}).

In Fig. \ref{fig:04:06} the time-varying PDP and DSD of our example measurement runs are depicted for $k_f=3$ (corresponding to the stationary frequency range of $5480-5520\,$MHz), and will be described next. 
The PDP is plotted in the upper row (Figs. (a) and (b)), and the DSD is shown in the lower row (Figs. (c) and (d)).
The plots on the left hand side of the figure correspond to the \emph{street crossing} scenario in an \emph{urban} environment. The ones on the right hand side correspond to measurements taken on the \emph{highway} with TX and RX driving in the same direction.
\vspace{0.1cm}
\begin{itemize}
\item \emph{Road crossing - urban single lane}\\
From $0$ to $7$ seconds the two vehicles approach the crossing from perpendicular streets without LOS, see Fig. \ref{fig:scenarios} (c). Due to multipath reflections on the buildings in the street or on other vehicles parked beside the street, the communication can be still established from the TX to the RX. Due to the higher attenuation, we observe a weak signal component during this time interval in Figs. \ref{fig:04:06} (a) and (c). 

Furthermore, in the DSD (Fig. \ref{fig:04:06} (c)) we notice that the signal component is curved, starting from values of a Doppler shift of $0$ Hz, increasing to $300$ Hz, and then decreasing again. This is caused by the TX vehicle, which was in a static position at the beginning of the measurement, accelerated at the start of the measurement, and braked when arriving at the crossing. We also observe some late components coming from reflections on other objects in the street for the PDP at $5$ s in Fig. \ref{fig:04:06} (a) and (c).
 
There is a clearly distinguishable region in Figs. \ref{fig:04:06} (a) and (c) between seconds $7$ and $10$. During this time interval both vehicles are at the crossing, the TX stops to give way to the RX, which passes by. Here, both a strong LOS component as well as more multipath components (MPCs) are present. This happens because the TX and RX are placed in a more open area where both the direct link and reflections with other near-by objects are stronger.

\vspace{0.1cm}
\item \emph{General LOS obstruction - highway}\\
The time-varying PDP and DSD for $40\,$MHz stationary frequency range are depicted in Figs. \ref{fig:04:06} (b) and (d).
At the beginning of the measurement, the direct path between the TX and the RX is obstructed by a truck, all three vehicles are driving at about $33.3\,$m/s ($120$\,km/h). During the measurement run, which lasts $10$ seconds, the truck in between the two vehicles \tz{changes} lanes and leaves the LOS free of obstruction.

Since this is a convoy measurement, the Doppler shift of the strongest component remains constant at $0\,$Hz in Fig. \ref{fig:04:06} (d), because the relative speed between TX and RX does not change. We can also appreciate that by looking at the PDP in Fig. \ref{fig:04:06} (b), where the first component remains also constant in delay. Furthermore, MPCs caused by reflections on vehicles and trucks driving in the same direction and in opposite direction are visible, mainly in the PDP.
\end{itemize}

\section{Time-Frequency-Varying RMS Delay Spread and RMS Doppler Spread}
\label{se:TVparams}
A radio channel can be described by its root mean square (RMS) delay spread and its RMS Doppler spread \cite{Molisch1999}, which have been usually assumed to be constant in time and frequency. However, as mentioned before, the fading process in vehicular channels is a non-stationary process with local stationarity within a finite stationarity region. Therefore, it makes sense to characterize the RMS delay spread and RMS Doppler spread as time-frequency-varying channel parameters. 

\subsection{Definition}
The second order central moments of the PDP and the DSD are important for the description of the fading process. They are directly related to the coherence bandwidth and coherence time of the channel, which indicate the rate of change of the channel in frequency and time, respectively \cite{Molisch1999}.

The time-frequency-varying RMS delay spread
\begin{equation}
\footnotesize
\sigma_{\tau}[k_t,k_f]=\sqrt{\frac{\sum\limits_{n=0}^{N-1}{(n\tau_\text{s})^2\hat{P}_{\tau}[k_t,k_f;n]}}
{\sum\limits_{n=0}^{N-1}{\hat{P}_{\tau}[k_t,k_f;n]}}-
\left(\frac{\sum\limits_{n=0}^{N-1}{(n\tau_\text{s})\hat{P}_{\tau}[k_t,k_f;n]}}{\sum\limits_{n=0}^{N-1}
{\hat{P}_{\tau}[k_t,k_f;n]}} \right)^2},
\end{equation}
and the time-frequency-varying RMS Doppler spread 
\begin{equation}
\footnotesize
\sigma_{\nu}[k_t,k_f]=\sqrt{\frac{\sum\limits_{p=-M/2}^{M/2-1}{(m\nu_\text{s})^2\hat{P}_{\nu}[k_t,k_f;p]}}
{\sum\limits_{p=-M/2}^{M/2-1}{\hat{P}_{\nu}[k_t,k_f;p]}}-
\left(\frac{\sum\limits_{p=-M/2}^{M/2-1}{(m\nu_\text{s})\hat{P}_{\nu}[k_t,k_f;p]}}{\sum\limits_{p=-M/2}^{M/2-1}{\hat{P}_{\nu}[k_t,k_f;p]}} \right)^2}
\end{equation}
are calculated using the estimated PDP ($\hat{P}_{\tau}$) and DSD ($\hat{P}_{\nu}$), respectively.  

Preprocessing of the estimated LSF is performed in order to eliminate spurious components that would lead to erroneous results. 
We apply a \emph{noise-power threshold} to eliminate noise components that could be mistaken as MPCs \cite{Molisch1999}, and it is chosen to be $5\,$dB above the noise floor that is determined from delay values larger than $2\,\mu$s. The thresholding is done separately for each stationarity region. On average the noise-power threshold is around $-85\,$\tz{dBm}.

Figure \ref{fig:TF_rms} shows the time-frequency-varying RMS delay spread for a \emph{general LOS obstruction} convoy measurement on the highway. Here, both the time variability and the frequency variability of this parameter are notable. For the sake of simplicity in the analysis that follows, we decide to select only one stationarity frequency region, namely the one corresponding to the frequency range $5480-5520\,$MHz, and look at the time-variability of the channel parameters.

\subsection{Empirical Results}
When analyzing the time-varying channel parameters, it is helpful to look at the time-varying PDP and DSD. In Fig. \ref{fig:04:06}, 
the presence of diffuse components is noteworthy, which is more significant in the \emph{urban} scenario. Furthermore, we observe some late components resulting from reflections on other objects beside the street.
We also highlight that the strong time-variability of the channel is more pronounced in the \emph{street crossing} measurement in the {urban} environment. 

Figure \ref{fig:04:07} depicts the obtained time-varying RMS delay spread and RMS Doppler spread for the two illustrative measurements for the frequency range $5480-5520\,$MHz. The results for the \emph{highway} measurement are plotted as gray solid line with star markers, and the results for the \emph{urban} scenario are shown as mere solid black line.
\vspace{0.1cm}
\begin{itemize}
\item \emph{RMS delay spread $\mathbf{\sigma_{\tau}}$}\\
In the \emph{general obstructed LOS} scenario, the RMS delay spread oscillates around $50$\,ns and it decreases when the late MPCs have no significant power in comparison to the strongest MPC, which happens between $1$ and $4$\,s, see Fig. \ref{fig:04:07} (a).

\begin{figure}
\begin{center}
\includegraphics[width=0.9\columnwidth]{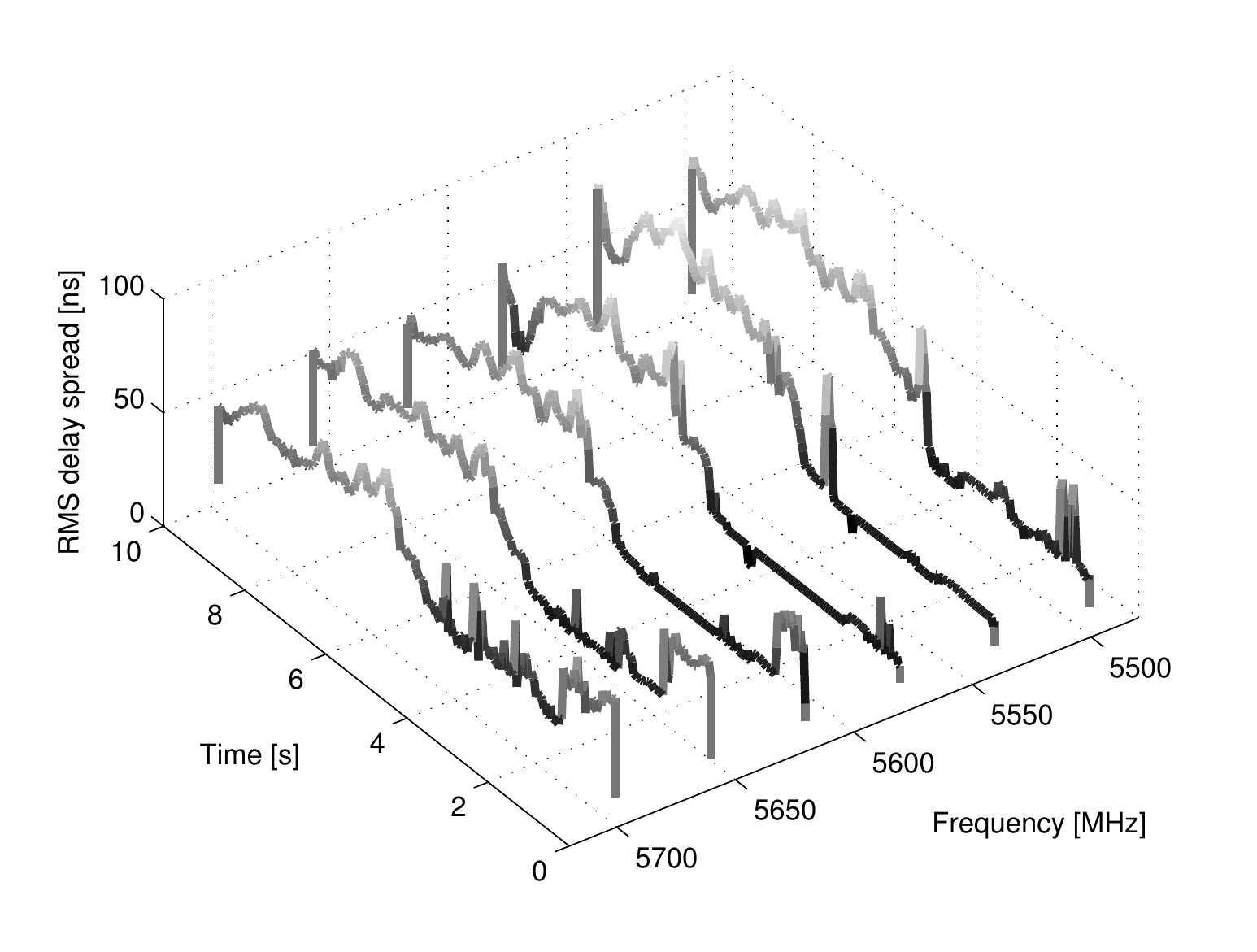}
\caption{Time-frequency-varying RMS delay spread for a \emph{general LOS obstruction} convoy measurement on the highway.}%
\label{fig:TF_rms}
\end{center}
\end{figure}

\begin{figure}[t!]
\begin{center}
\subfigure[time-varying RMS delay spread]{\includegraphics[width=0.9\columnwidth]{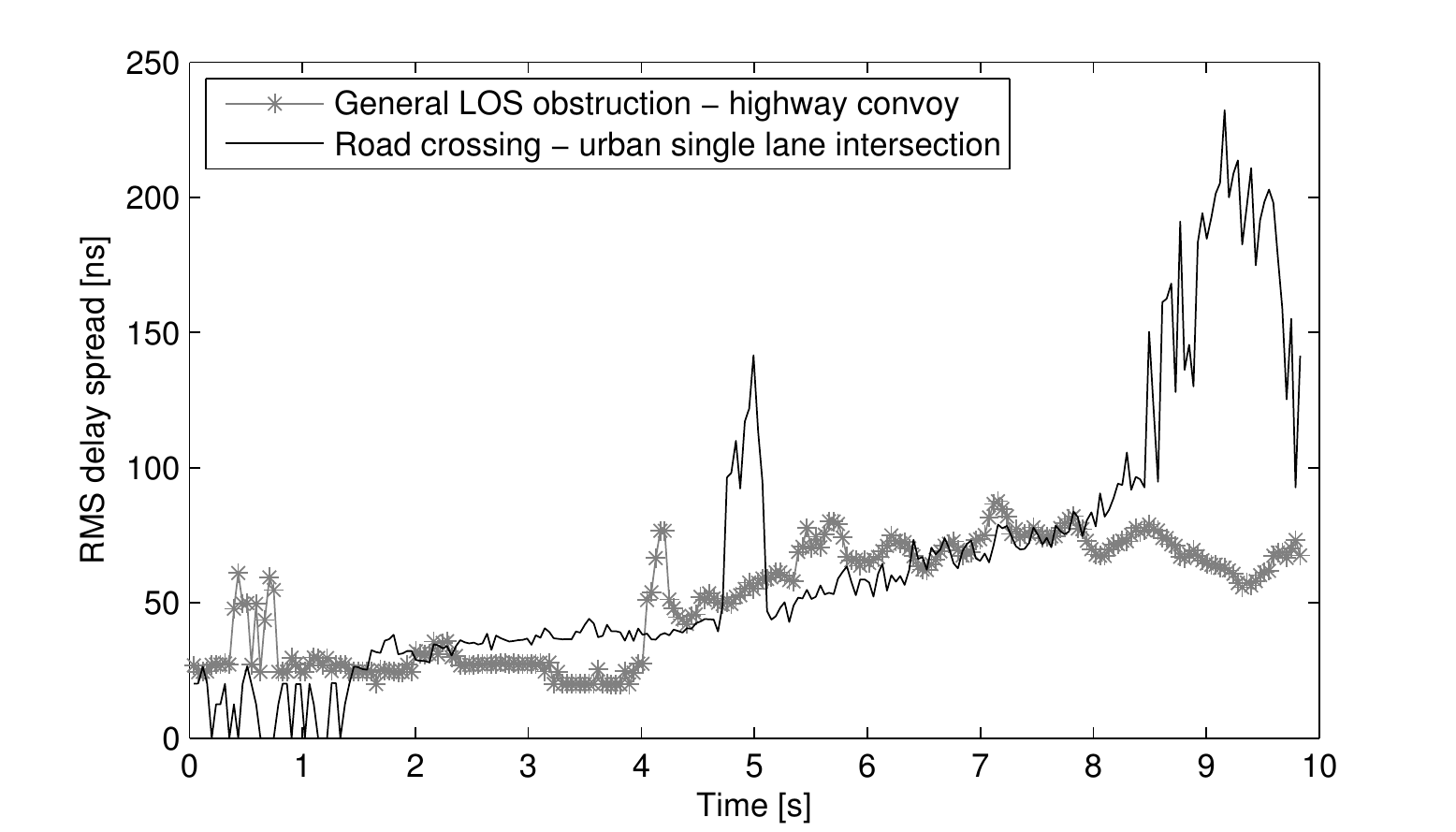}}%
\\
\subfigure[time-varying RMS Doppler spread]{\includegraphics[width=0.9\columnwidth]{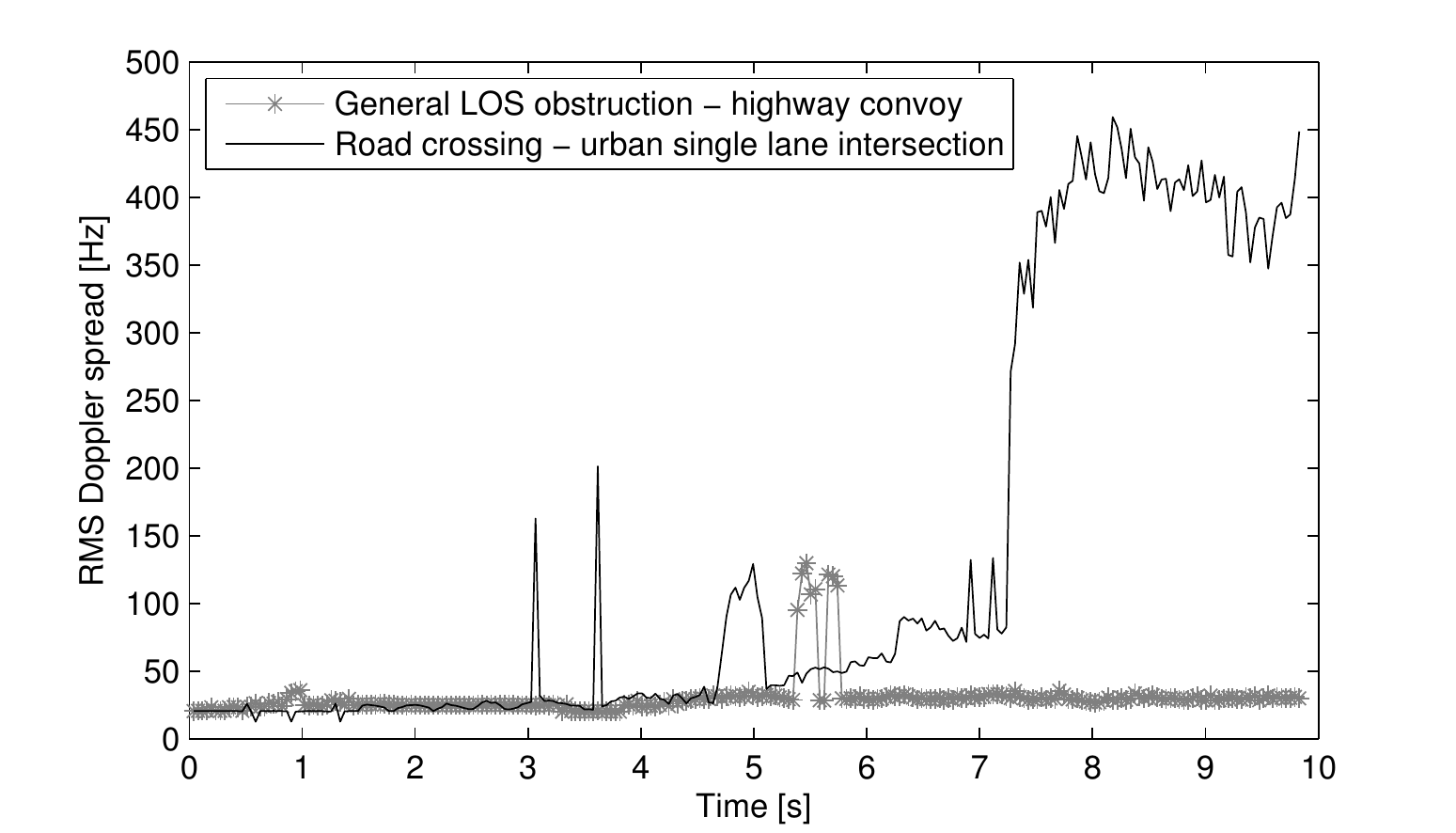}}%
\caption{Time-varying channel parameters for the frequency range $5480$-$5520\,$MHz for two different measurement scenarios.}%
\label{fig:04:07}
\end{center}
\end{figure}

For the \emph{street crossing} scenario, the RMS delay spread is higher towards the end of the measurement run due to the richness of MPCs present in the environment at this moment. Note the peak right before $5\,$s, which occurs due to a strong late contribution, as observable in the PDP in Fig. \ref{fig:04:06} (a). 

\vspace{0.1cm}
\item \emph{RMS Doppler spread $\mathbf{\sigma_{\nu}}$}\\
There is a big difference in the RMS Doppler spread for the two selected measurements, as one can clearly see in Fig. \ref{fig:04:07} (b). The \emph{general LOS obstruction} scenario corresponds to a convoy measurement, i.e. the two vehicles drive in the same direction at more or less constant speed. This results in a constant Doppler component at $0\,$Hz, see Fig. \ref{fig:04:06} (d). The RMS Doppler spread for this case remains also roughly constant at around $33\,$Hz. Nevertheless, it can happen that the RMS Doppler spread increases when new MPCs with strong power are present, as in Fig. \ref{fig:04:06} (d) between $5$ and $6\,$s.

On the other hand, the \emph{street crossing} measurement consists of two vehicles approaching each other, i.e. an oncoming measurement. Here we observe a relatively constant RMS Doppler spread around $20$ Hz until $4.5\,$s. As we can see in the DSD plot in Fig. \ref{fig:04:06} (b), we only have one strong Doppler component from $4.5\,$s until $5\,$s, which contributes to increase the RMS Doppler spread. From $5\,$s and onwards, the strong MPC disappears and more Doppler components are present which increases the RMS Doppler spread. This results in the increase of the RMS Doppler spread at $7.3$ s. At that point the RMS Doppler spread reaches a maximum value and oscillates around $400$ Hz. The number of different Doppler components relevant during this period remains also fairly large, as can be observed in Fig. \ref{fig:04:06} (c).
\end{itemize}

\section{Statistical Modeling}
\label{se:statistics}
Until now we have analyzed single measurement runs for two different scenarios for one stationarity region in frequency ($40\,$MHz), but actually we want to characterize the entire measurement data set and derive a meaningful statistical parametrization for each scenario. Therefore, we perform the same analysis for each individual measurement run for all scenarios and for all stationarity regions in frequency ($6$ regions of $40\,$MHz each over the $240\,$MHz of total measurement bandwidth).

\subsection{Fitting of Empirical Distributions}
In order to smoothly proceed from the results presented previously, we continue with the selected example scenarios for our illustrations. \tz{Based on the shape of the histograms (not shown), we chose a bi-modal Gaussian mixture distribution for statistical characterization. The probability density function (PDF) $p$ of a bi-modal Gaussian mixture distribution reads \cite{McLachlan2000}
\begin{equation}
f(z)=\frac{1}{\alpha}\left(\frac{w}{\sqrt{2\pi}\,\sigma_1}e^{-\frac{(z-\mu_1)^2}{2\sigma_1}} + \frac{1-w}{\sqrt{2\pi}\,\sigma_2}e^{-\frac{(z-\mu_2)^2}{2\sigma_2}} \right),
\end{equation}
where $z$ denotes the random variable to be described by the PDF, i.e. either the RMS delay spread $\sigma_\tau$ or the RMS Doppler spread $\sigma_\nu$. The bi-modal Gaussian mixture is truncated to the interval $z\in(0,z_\text{max})$ were $z_\text{max}$ either represents the maximum RMS delay spread $\sigma_{\tau,\text{max}}$ or the maximum RMD Doppler spread $\sigma_{\nu,\text{max}}$. To obtain a true PDF we correct for the truncated tail of the PDF with the coefficient $\alpha=F'_0(z_\text{max})- F'_0(0)$ with
\begin{equation}
F'_0(z)= w\left(1 - Q\left(\frac{z-\mu_1}{\sigma_1}\right) \right) + (1-w)\left(1 - Q\left(\frac{z-\mu_2}{\sigma_2}\right) \right)
\end{equation}
denoting the cumulative distribution function (CDF) of the bi-model Gaussian mixture without truncation and  
\begin{equation}
F_0(z)=\alpha (F'_0(z) - F'_0(0))
\end{equation}
for $z\in(0,z_\text{max})$ denotes the CDF of the truncated bi-modal distribution. The parameters listed in Tab. \ref{tab:04:08} are the mean $\mu_x$ and standard deviation $\sigma_x$, where $x\in\{1,2\}$ indexes each Gaussian component, and $w$ is the weighting factor.}

\begin{figure}
\begin{center}
\subfigure[CDF RMS delay spread]{\includegraphics[width=0.9\columnwidth]{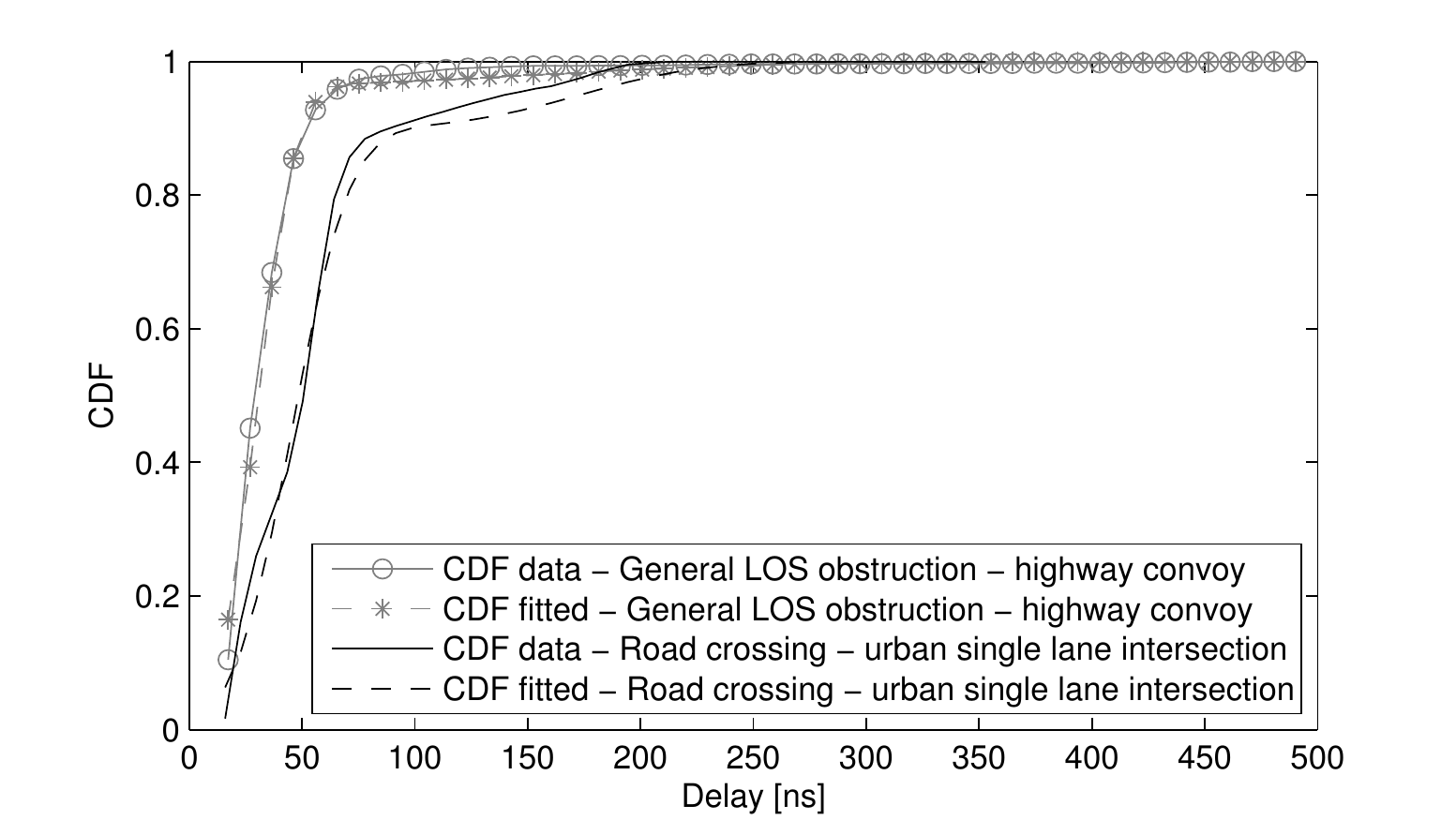}}
\subfigure[CDF RMS Doppler spread]{\includegraphics[width=0.9\columnwidth]{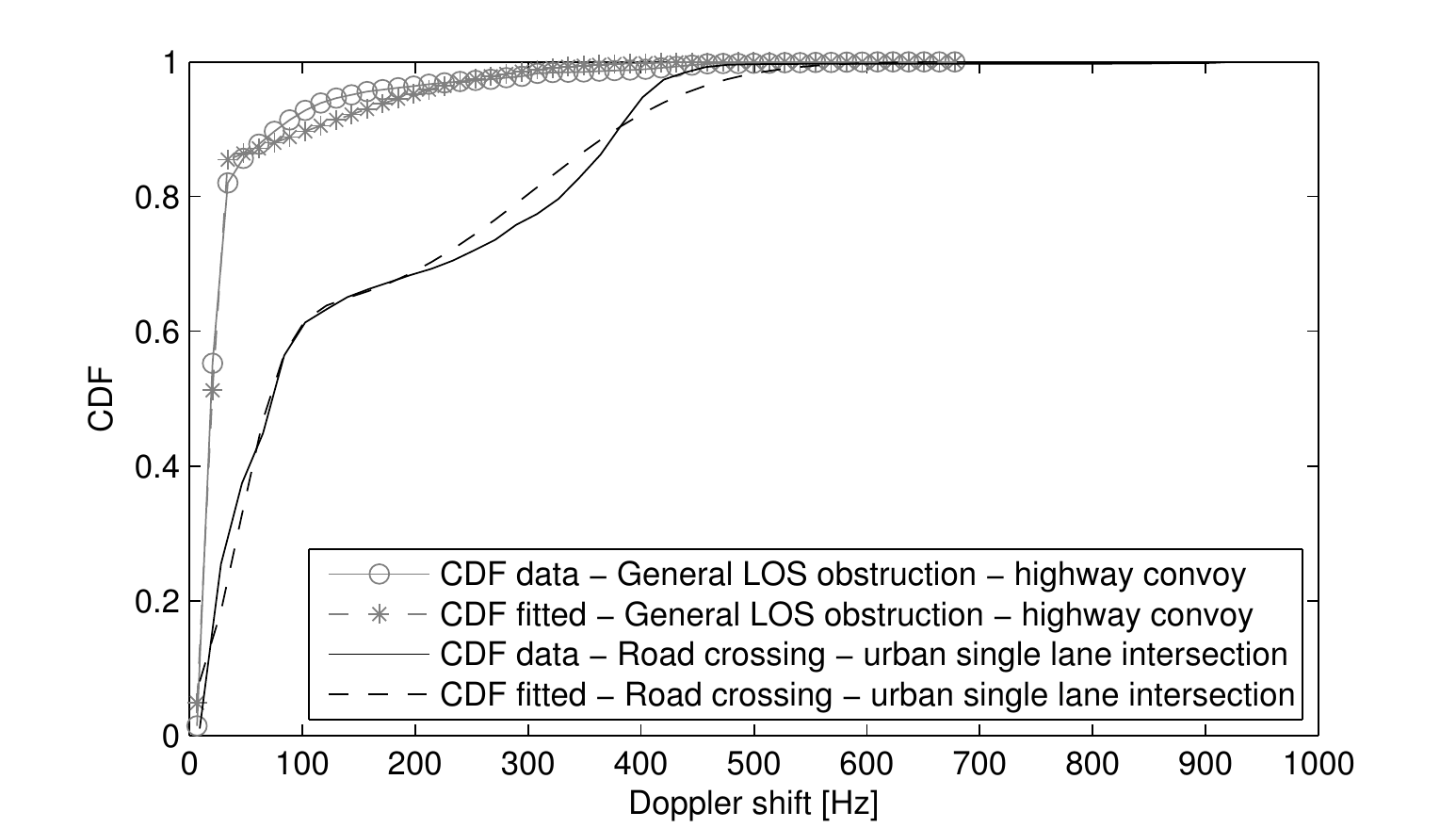}}
\caption{CDF of the time-frequency-varying channel parameters for the illustrative scenarios. The empirical CDF is plotted in solid line, and the fitted CDF in dashed lines.}
\label{fig:04:08}
\end{center}
\end{figure}

We plot in Fig. \ref{fig:04:08} the cumulative distribution function (CDF) of the time-frequency-varying channel parameters for all measurement runs performed under the same conditions. The solid lines correspond to the empirical CDF obtained from the measurements, in dashed lines we plot the fitted CDF. The two different scenarios are distinguished by the marker symbols: gray lines with star markers correspond to the \emph{general LOS obstruction} scenario, whereas the black lines without markers are employed for the \emph{street crossing} scenario.

We show the agreement between the empirical CDF $F_Z(z)$ and the analytical CDF $F_0(z)$ in Fig. \ref{fig:04:08}. Moreover, we use here also the Kolmogorov-Smirnov (KS) test as a goodness-of-fit (GoF) indicator \cite{Massey1951} calculating
\begin{equation}
\text{GoF} =\sup_z|F_Z(z)-F_0(z)|\,,
\end{equation}
where $\sup$ denotes the supremum. The result of this test is listed in Tab. \ref{tab:04:08} in the column GoF. The significance level of $\text{GoF}\le\gamma=0.11$ indicates a good fit. 

A bi-modal Gaussian mixture distribution is adequate for modeling these channel parameters since the channel parameters are highly dependent on the presence, or lack, of strong MPCs. The first Gaussian components encompasses the channel parameters values for a LOS situation when later MPCs have minor relevance. On the other hand, for non-LOS situations where the later MPCs are significant, the channel parameter values are described by the second Gaussian component.

We perform the same analysis for the whole data set and depict the fitting parameters in Tab. \ref{tab:04:08}. Looking at the weighting \tz{factors $w$ and $1-w$, we observe} that the weighting factor for one of the Gaussian components is always much greater than for the other component. During the measurements there is a transition between LOS and non-LOS situations, the weighting factor indicates how often these two situations are observed in each scenario. When $w$ is close to $1$ or $0$ the probability of transition will be low. 

\tz{Please note that the mean values $\mu_1$ and $\mu_2$ of the bi-modal Gaussian mixture can be interpreted as the mean values of the channel parameter (i.e. the RMS delay or RMS Doppler spread) for LOS and non-LOS situations within a single scenario. While the standard deviation is a measure of the non-stationarity of the channel. The larger the standard deviation is, the less stationary is the channel and the more the channel parameter (i.e. the RMS delay or RMS Doppler spread) will vary within one scenario\footnote{This interpretation must be handled with care for some scenarios, since the bi-modal Gaussian mixture is truncated at $0$ and at the maximum observed RMS delay spread or RMS Doppler spread. E.g. for the scenario \emph{street crossing - suburban with traffic} the high value of $\sigma_2$ for the RMS Doppler spread indicates that the PDF has one pronounced peak at the value of $\mu_1$ and a flat shape around the value of $\mu_2$.}.}

\begin{table*}
\caption{Modeling time-frequency-varying channel parameters}
\label{tab:04:08}
\begin{center}
 \begin{tabular}{@{}llcccccccccc@{}} \toprule
  && \tz{$w$}  & $\mu_1$ & $\sigma_1$ & $\mu_2$ & $\sigma_2$ & GoF & max. & $B_{\mathrm{coh}}$ [kHz] & $T_{\mathrm{coh}}$ [$\mu$s] & runs\\
\midrule

\multicolumn{10}{@{}l@{}}{\emph{\textbf{street crossing - suburban with traffic}}}\\\addlinespace[1mm]

$\mathbf{\sigma_{\tau}}$ \textbf{[ns]} && $0.65$ & $47.07$ & $13.72$ 	& $127.88$ & $50.99$ & $<0.02$ &$255.77$ & $651.62$ &- & $3$\\
$\mathbf{\sigma_{\nu}}$ \textbf{[Hz]} && $0.32$ & $39.55$ &$101.69$ & $156.62$ & $9060.7$ & $<0.05$ &$352.93$ &- & $472.23$ & $3$\\\addlinespace[1mm]
\midrule

\multicolumn{10}{@{}l@{}}{\emph{\textbf{street crossing - suburban without traffic}}}\\addlinespace[1mm]
$\mathbf{\sigma_{\tau}}$ \textbf{[ns]} && $0.69$ & $43.87$ & $14.14$ 	& $167.83$ &$51.73$ & $<0.02$&$808.23$ & $206.21$ &- & $11$\\
$\mathbf{\sigma_{\nu}}$ \textbf{[Hz]} && $0.45$ & $29.59$ &$14.58$ 	& $211.08$ & $86.49$ & $<0.03$ &$684.03$ &- & $243.65$ & $11$\\\addlinespace[1mm]
\midrule

\multicolumn{10}{@{}l@{}}{\emph{\textbf{street crossing - urban single lane}}}\\\addlinespace[1mm]
$\mathbf{\sigma_{\tau}}$ \textbf{[ns]} &&  $0.90$  & $45.57$ & $20.07$ & $173.13$ & $42.72$ & $<0.07$&$925.66$ & $180.05$ &- & $5$\\
$\mathbf{\sigma_{\nu}}$ \textbf{[Hz]} &&  $0.62$ & $45.05$ & $31.21$ & $306.04$ & $114.94$ & $<0.07$ &$933.70$ &- & $178.50$ & $5$\\\addlinespace[1mm]
\midrule

\multicolumn{10}{@{}l@{}}{\emph{\textbf{street crossing - urban multiple lane}}}\\\addlinespace[1mm]
$\mathbf{\sigma_{\tau}}$ \textbf{[ns]} && $1$ & $50.24$ & $24.81$ & $-$ & $-$ & $<0.02$&$926.66$ &$179.86$ &- & $5$\\
$\mathbf{\sigma_{\nu}}$ \textbf{[Hz]} && $0.88$ & $35.48$ & $24.40$ & $274.86$	& $144.12$ & $<0.11$ &$822.75$ &- & $202.57$ & $5$\\\addlinespace[1mm]
\midrule

\multicolumn{10}{@{}l@{}}{\emph{\textbf{general LOS obstruction - highway}}}\\\addlinespace[1mm]
$\mathbf{\sigma_{\tau}}$ \textbf{[ns]} && $0.95$ & $30.12$ & $13.09$ & $153.50$ &  $114.11$ & $<0.06$&$674.95$ &$246.93$ &- & $12$\\
$\mathbf{\sigma_{\nu}}$ \textbf{[Hz]} && $0.80$ & $19.52$ & $5.15$ 	& $108.13$ &  $128.18$ & $<0.04$ &$684.83$ &- & $243.37$ & $12$\\\addlinespace[1mm]
\midrule

\multicolumn{10}{@{}l@{}}{\emph{\textbf{merging lanes - rural}}}\\\addlinespace[1mm]
$\mathbf{\sigma_{\tau}}$ \textbf{[ns]} && $0.86$ & $29.10$ &  $7.44$ & $173.77$ & $66.78$ & $<0.05$&$254.45$ &$655.02$ &- & $7$\\
$\mathbf{\sigma_{\nu}}$ \textbf{[Hz]} && $0.45$ & $22.29$ & $3.92$ & $100.78$ & $87.32$ & $<0.06$ &$402.61$ &- & $413.97$ & $7$\\\addlinespace[1mm]
\midrule

\multicolumn{10}{@{}l@{}}{\emph{\textbf{traffic congestion - slow traffic}}}\\\addlinespace[1mm]
$\mathbf{\sigma_{\tau}}$ \textbf{[ns]} && $0.87$ & $30.40$ & $11.46$ & $139.95$ & $54.85$ & $<0.07$&$924.79$ & $180.22$ &- & $11$\\
$\mathbf{\sigma_{\nu}}$ \textbf{[Hz]} && $0.73$ & $20.05$ & $6.02$ & $113.90$ & $73.28$ & $<0.03$ &$849.91$ &- & $196.10$ & $11$\\\addlinespace[1mm]
\midrule

\multicolumn{10}{@{}l@{}}{\emph{\textbf{traffic congestion - approaching traffic jam}}}\\\addlinespace[1mm]
$\mathbf{\sigma_{\tau}}$ \textbf{[ns]} && $0.88$ & $28.99$ & $9.17$ & $136.53$ & $63.46$ & $<0.04$&$677.20$ & $246.11$ &- & $7$\\
$\mathbf{\sigma_{\nu}}$ \textbf{[Hz]} && $0.64$ & $23.49$ & $4.30$ & $74.66$ & $58.34$ & $<0.05$ &$511.78$ &- & $325.66$ & $7$\\\addlinespace[1mm]
\midrule

\multicolumn{10}{@{}l@{}}{\emph{\textbf{in-tunnel}}}\\\addlinespace[1mm]
$\mathbf{\sigma_{\tau}}$ \textbf{[ns]} && $0.95$ & $75.11$ & $23.99$ & $109.77$ & $43.44$ & $<0.04$&$244.75$ & $680.98$ &- & $7$\\
$\mathbf{\sigma_{\nu}}$ \textbf{[Hz]} && $0.77$ & $89.56$ & $51.08$ & $159.24$	& $84.31$ & $<0.04$ &$492.56$ &- & $338.37$ & $7$\\\addlinespace[1mm]
\midrule

\multicolumn{10}{@{}l@{}}{\emph{\textbf{on-bridge}}}\\\addlinespace[1mm]
$\mathbf{\sigma_{\tau}}$ \textbf{[ns]} && $0.78$ & $38.24$ & $12.94$& $167.48$& $98.85$ & $<0.03$&$951.07$ & $175.24$ &- & $4$\\
$\mathbf{\sigma_{\nu}}$ \textbf{[Hz]} && $0.69$ & $81.71$ & $37.94$ & $163.19$	& $100.40$ & $<0.06$ &$895.48$ &- & $186.12$ & $4$\\
\bottomrule
\end{tabular}
\end{center}
\end{table*}

\subsection{Discussion of Statistical Results}
After processing all measurement runs we can draw conclusions based on the obtained statistical distributions for each traffic-scenario. It is important to mention that the line of argument is based on mean values. It can however happen that at a given time instant, values of RMS delay and Doppler spread do not follow the described trend. Critical values of the RMS spreads for a communication system are going to be the extreme maxima. We display these extreme maxima in the fourth last column of Tab. \ref{tab:04:08}.

\subsubsection{RMS delay spread $\mathbf{\sigma_{\tau}}$} 
Interestingly, low RMS delay spreads are obtained in \emph{highway} environments, in the \emph{traffic congestion} scenario with mean values around $30\,$ns. As it was already observed in \cite{Paier2010}, other vehicles driving beside the TX and the RX are scatterers that are not as relevant as one might expect. This is due to the fact that the roof-top antenna in our set-up is placed at a position slightly above the other vehicles and that the antenna patterns show a low gain in the horizontal or below horizontal elevation angle. More relevant MPCs stem from big scattering objects such as trucks or big metallic structures. 

Slightly higher RMS mean delay spreads have been obtained in \emph{street crossing} situations, with values between $40$ and $50\,$ns. 
The highest mean RMS delay spread values is obtained for the \emph{in-tunnel} scenario, 
where big metallic structures, e.g. the ventilation system in the tunnel, 
placed relatively close to the vehicles, contribute to increasing the RMS delay spread.

The two merging streets in the \emph{merging lane} scenario were located in a \emph{rural} environment, with very few scattering objects in the surroundings and not much traffic, therefore the RMS delay spread is small.

During the measurements taken for the \emph{general LOS obstruction} scenario, several trucks were driving beside the TX and the RX and were blocking the LOS. The RMS delay spread in this scenario is mainly determined by the presence of these big objects.
The maximum RMS delay spread values are below $1\,\mu$s, ranging from $200$ to $900\,$ns.

\subsubsection{RMS Doppler spread $\mathbf{\sigma_{\nu}}$}
In the case of RMS Doppler spread, the weighting factors vary depending on the relative speed between TX and RX, and the angle between them. 

The RMS Doppler spread is in general larger for the \emph{street crossing} scenarios. A reason for that is the rapid change of the main Doppler component from positive to negative values (approaching and leaving). Also high RMS Doppler spreads are observed in the \emph{in-tunnel} and \emph{on-bridge} scenarios. There, the later arriving MPCs caused by reflections on metallic surfaces are strong and therefore contribute to enlarging the RMS Doppler spread values.

The RMS Doppler spreads tend to remain constant at low levels in scenarios where the TX and RX vehicles are driving in the \emph{same direction}, at the same speed, and where the MPCs are not strong. This is somehow expected since the most relevant Doppler component remains around $0$ Hz throughout all the measurement runs for same direction measurements.

The maximum RMS Doppler spread values are in the order of hundreds of Hz, mostly between $400$ and $600\,$Hz with a maximum of $933.70\,$Hz.


\subsubsection{Coherence Bandwidth and Coherence Time}
A further insight about the frequency selectivity of the channel can be derived from these channels parameters and be used in system performance simulations. The RMS delay spread relates to the coherence bandwidth through an uncertainty relationship as $B_\text{coh,k}\ge\arccos(k)/ 2\pi\sigma_{\tau}$, with $k$ being a specific level such that the channel autocorrelation function $\mid R_H(t,B_{\text{coh,k}}) \mid\ < k$ \cite{Fleury1996}. The coherence bandwidth indicates the severity of the channels' frequency selectivity. The time selectivity is reflected in the RMS Doppler spread and relates to the coherence time as $T_\text{coh,k}\ge\arccos(k)/2\pi\sigma_{\nu}$, with $k$ such that $\mid R_H(T_{\text{coh,k}},f) \mid\ < k$. Usual values of $k$ are $0.5$ and $0.75$ \cite{Molisch1999}. \tz{We choose $k=0.5$ for calculating the coherence parameters, with $T_\text{coh}$ computed form the maximum RMS Doppler spread, and $B_\text{coh}$ computed from the maximum RMS delay spread. The results are listed as extra columns on the right side in Tab. \ref{tab:04:08}.} Coherence times are in the range from $180$ to $500\,\mu$s, and coherence bandwidths range from $200\,$kHz to $700\,$kHz. Hence the vehicular channel in all measured scenarios is strongly time- and frequency-selective.

\section{Conclusions}
\label{se:conclusions}
In this paper the non-stationary fading process of vehicular channels is analyzed. For this purpose we used radio channel measurement data collected in safety-relevant traffic scenarios for intelligent transportation systems (ITS). 
 
The local scattering function (LSF) is estimated based on the measured sampled time-varying frequency response. We derived the time-frequency-varying projections of the LSF on the delay and the Doppler domain, to obtain the time-frequency-varying power delay profile (PDP), and the time-frequency-varying Doppler power spectral density (DSD) of the non-stationary fading process. Based on these results we calculated the second order central moment of the PDP and DSD defining the time-frequency-varying root mean square (RMS) delay spread and the RMS Doppler spread. The empirical distribution of the RMS delay spread and Doppler spread for each traffic scenario was fitted by means of a simple but accurate bi-modal Gaussian model.

By looking at the mean values of the time-frequency-varying parameters for all measured scenarios, we conclude that high RMS delay spreads are mostly observed: 
\emph{(i)} in situations with obstruction of the line of sight with big reflecting objects driving beside the TX and RX direct link, and \emph{(ii)} in environments where big, often metallic, structures are placed close to the TX-RX.
High RMS Doppler spreads occur in: \emph{(i)} drive-by scenarios, and \emph{(ii)} situations where late Doppler components are significant, mainly caused by well-reflecting objects.

\section*{Acknowledgement}
We would like to thank all the participants who made possible the conduction of the DRIVEWAY'09 measurement campaing: Nicolai Czink, Johan Karedal, Oliver Klemp, Andreas Kwoczek, Alexander Paier, Andreas Thiel, Yi Zhou.

\bibliography{IEEEfull,biblio_diss}

\begin{thebibliography}{10}
\providecommand{\url}[1]{#1}
\csname url@samestyle\endcsname
\providecommand{\newblock}{\relax}
\providecommand{\bibinfo}[2]{#2}
\providecommand{\BIBentrySTDinterwordspacing}{\spaceskip=0pt\relax}
\providecommand{\BIBentryALTinterwordstretchfactor}{4}
\providecommand{\BIBentryALTinterwordspacing}{\spaceskip=\fontdimen2\font plus
\BIBentryALTinterwordstretchfactor\fontdimen3\font minus
  \fontdimen4\font\relax}
\providecommand{\BIBforeignlanguage}[2]{{%
\expandafter\ifx\csname l@#1\endcsname\relax
\typeout{** WARNING: IEEEtran.bst: No hyphenation pattern has been}%
\typeout{** loaded for the language `#1'. Using the pattern for}%
\typeout{** the default language instead.}%
\else
\language=\csname l@#1\endcsname
\fi
#2}}
\providecommand{\BIBdecl}{\relax}
\BIBdecl

\bibitem{WAVEStandard}
``IEEE P802.11p: Part 11: Wireless LAN Medium Access Control (MAC) and Physical
  Layer (PHY) Specifications: Amendment 6: Wireless Access in Vehicular
  Environments,'' July 2010.

\bibitem{Ivan2009}
I.~Ivan, P.~Besnier, M.~Crussiere, M.~Drissi, L.~Le~Danvic, M.~Huard, and
  E.~Lardjane, ``Physical layer performance analysis of {V2V} communications in
  high velocity context,'' in \emph{9th International Conference on Intelligent
  Transport Systems Telecommunications (ITST)}, October 2009, pp. 409--414.

\bibitem{Lin2011}
C.-S. Lin, C.-K. Sun, J.-C. Lin, and B.-C. Chen, ``Performance evaluations of
  channel estimations in {IEEE} 802.11p environments,'' \emph{Telecommunication
  Systems}, 2011.

\bibitem{Kiokes2009}
G.~Kiokes, A.~Amditis, and N.~Uzunoglu, ``Simulation-based performance analysis
  and improvement of orthogonal frequency division multiplexing - 802.11p
  system for vehicular communications,'' \emph{IET Intelligent Transport
  Systems}, vol.~3, no.~4, pp. 429--436, December 2009.

\bibitem{Nuckelt2011}
J.~Nuckelt, M.~Schack, and T.~K\"urner, ``{Deterministic and stochastic channel
  models implemented in a physical layer simulator for Car-to-X
  communications},'' \emph{Advances in Radio Science}, vol.~9, pp. 165--171,
  2011.

\bibitem{Reichardt2011b}
L.~Reichardt, L.~Sit, T.~Schipper, and T.~Zwick, ``{IEEE} 802.11p based
  physical layer simulator for car-to-car communication,'' in \emph{5th
  European Conference on Antennas and Propagation (EUCAP)}, April 2011, pp.
  2876--2880.

\bibitem{Bernado2010a}
L.~Bernad\'o, T.~Zemen, N.~Czink, and P.~Belanovi\'c, ``Physical layer
  simulation results for {IEEE} 802.11p using vehicular non-stationary channel
  model,'' in \emph{IEEE International Conference on Communications Workshops
  (ICC)}, May 2010.

\bibitem{Zemen2012}
T.~Zemen, L.~Bernad\'o, N.~Czink, and A.~F. Molisch, ``Iterative time-variant
  channel estimation for 802.11p using generalized discrete prolate spheroidal
  sequences,'' \emph{IEEE Transactions on Vehicular Technology}, vol.~61,
  no.~3, pp. 1222--1233, March 2012.

\bibitem{Zemen12a}
T.~Zemen and A.~F. Molisch, ``Adaptive reduced-rank estimation of
  non-stationary time-variant channels using subspace selection,'' \emph{{IEEE}
  Transactions on Vehicular Technology}, vol.~61, no.~9, pp. 4042--4056,
  November 2012.

\bibitem{Fernandez12}
J.~Fernandez, K.~Borries, L.~Cheng, V.~Bhagavatula, D.~Stancil, and F.~Bai,
  ``Performance of the 802.11p physical layer in vehicle-to-vehicle
  environments,'' \emph{{IEEE} Transactions on Vehicular Technology}, vol.~61,
  no.~1, pp. 3--14, January 2012.

\bibitem{Matz2005}
G.~Matz, ``On non-{WSSUS} wireless fading channels,'' \emph{{IEEE} Transactions
  on Wireless Communications}, vol.~4, no.~5, pp. 2465--2478, September 2005.

\bibitem{Wilink2008}
T.~Willink, ``Wide-sense stationarity of mobile {MIMO} radio channels,''
  \emph{{IEEE} Transactions on Vehicular Communications}, vol.~57, no.~2, pp.
  704--714, March 2008.

\bibitem{Bernado2008}
L.~Bernad\'o, T.~Zemen, A.~Paier, G.~Matz, J.~Karedal, N.~Czink, F.~Tufvesson,
  M.~Hagenauer, A.~F. Molisch, and C.~F. Mecklenbr\"auker, ``{Non-WSSUS
  Vehicular Channel Characterization at 5.2 {GHz} - Spectral Divergence and
  Time-Variant Coherence Parameters},'' in \emph{Assembly of the International
  Union of Radio Science (URSI)}, August 2008, pp. 9--15.

\bibitem{Renaudin2010}
O.~Renaudin, V.-M. Kolmonen, P.~Vainikainen, and C.~Oestges, ``Non-stationary
  narrowband {MIMO} inter-vehicle channel characterization in the 5-{GHz}
  band,'' \emph{{IEEE} Transactions on Vehicular Technology}, vol.~59, no.~4,
  pp. 2007--2015, May 2010.

\bibitem{Mecklenbraeuker2011}
C.~F. Mecklenbr\"auker, A.~F. Molisch, J.~Karedal, F.~Tufvesson, A.~Paier,
  L.~Bernad\'o, T.~Zemen, O.~Klemp, and N.~Czink, ``Vehicular channel
  characterization and its implications for wireless system design and
  performance,'' \emph{Proceedings of the IEEE}, vol.~99, no.~7, pp.
  1189--1212, July 2011.

\bibitem{Molisch2009}
A.~F. Molisch, F.~Tufvesson, J.~Karedal, and C.~F. Mecklenbrauker, ``A survey
  on vehicle-to-vehicle propagation channels,'' \emph{{IEEE} Wireless
  Communications Magazine}, vol.~16, no.~6, pp. 12--22, December 2009.

\bibitem{Maurer2002}
J.~Maurer, T.~Fugen, and W.~Wiesbeck, ``Narrow-band measurement and analysis of
  the inter-vehicle transmission channel at 5.2 {GHz},'' in \emph{IEEE 55th
  Vehicular Technology Conference (VTC Spring)}, vol.~3, 2002, pp. 1274--1278.

\bibitem{Acosta2004}
G.~Acosta, K.~Tokuda, and M.~Ingram, ``Measured joint {Doppler}-delay power
  profiles for vehicle-to-vehicle communications at 2.4 {GHz},'' in \emph{IEEE
  Global Telecommunications Conference (GLOBECOM)}, vol.~6, December 2004, pp.
  3813--3817.

\bibitem{Paier2007}
A.~Paier, J.~Karedal, N.~Czink, H.~Hofstetter, C.~Dumard, T.~Zemen,
  F.~Tufvesson, A.~F. Molisch, and C.~F. Mecklenbr\"auker, ``Car-to-car radio
  channel measurements at 5 {GHz}: Pathloss, power-delay profile, and
  delay-{Doppler} spectrum,'' in \emph{4th International Symposium on Wireless
  Communication Systems (ISWCS)}, October 2007, pp. 224--228.

\bibitem{Karedal2011}
J.~Karedal, N.~Czink, A.~Paier, F.~Tufvesson, and A.~F. Molisch, ``Path loss
  modeling for vehicle-to-vehicle communications,'' \emph{{IEEE} Transactions
  on Vehicular Technology}, vol.~60, no.~1, pp. 323--328, January 2011.

\bibitem{Paschalidis2011}
P.~Paschalidis, K.~Mahler, A.~Kortke, M.~Peter, and W.~Keusgen, ``Pathloss and
  multipath power decay of the wideband car-to-car channel at 5.7 {GHz},'' in
  \emph{73rd IEEE Vehicular Technology Conference (VTC Spring)}, May 2011.

\bibitem{Kunisch2008}
J.~Kunisch and J.~Pamp, ``Wideband car-to-car radio channel measurements and
  model at 5.9 {GHz},'' in \emph{68th IEEE Vehicular Technology Conference (VTC
  Fall)}, September 2008.

\bibitem{Renaudin2008}
O.~Renaudin, V.-M. Kolmonen, P.~Vainikainen, and C.~Oestges, ``Wideband {MIMO}
  car-to-car radio channel measurements at 5.3 {GHz},'' in \emph{68th IEEE
  Vehicular Technology Conference (VTC Fall)}, September 2008.

\bibitem{Tan2008}
I.~Tan, W.~Tang, K.~Laberteaux, and A.~Bahai, ``Measurement and analysis of
  wireless channel impairments in {DSRC} vehicular communications,'' in
  \emph{IEEE International Conference on Communications (ICC)}, May 2008, pp.
  4882--4888.

\bibitem{Sen2008}
I.~Sen and D.~Matolak, ``Vehicle-vehicle channel models for the 5-{GHz} band,''
  \emph{{IEEE} Transactions on Intelligent Transportation Systems}, vol.~9,
  no.~2, pp. 235--245, June 2008.

\bibitem{Matz2003a}
G.~Matz, ``Doubly underspread non-{WSSUS} channels: Analysis and estimation of
  channel statistics,'' in \emph{4th IEEE Workshop on Signal Processing
  Advances in Wireless Communications (SPAWC)}, Rome, Italy, June 2003, pp.
  190--194.

\bibitem{Paier2008}
A.~Paier, T.~Zemen, L.~Bernad\'o, G.~Matz, J.~Karedal, N.~Czink, C.~Dumard,
  F.~Tufvesson, A.~F. Molisch, and C.~F. Mecklenbr\"auker, ``{Non-WSSUS
  vehicular channel characterization in highway and urban scenarios at 5.2
  {GHz} using the local scattering function},'' in \emph{International ITG
  Workshop on Smart Antennas (WSA)}, February 2008, pp. 9--15.

\bibitem{Okonkwo2010}
U.~Okonkwo, S.~Hashim, R.~Ngah, N.~Nanyan, and T.~Rahman, ``Time-scale domain
  characterization of nonstationary wideband vehicle-to-vehicle propagation
  channel,'' in \emph{IEEE Asia-Pacific Conference on Applied Electromagnetics
  (APACE)}, November 2010.

\bibitem{Chelli2011}
A.~Chelli and M.~P\"atzold, ``A non-stationary {MIMO} vehicle-to-vehicle
  channel model derived from the geometrical street model,'' in \emph{IEEE
  Vehicular Technology Conference (VTC Fall)}, September 2011.

\bibitem{Bernado2011b}
L.~Bernad\'o, A.~Roma, A.~Paier, T.~Zemen, N.~Czink, J.~Karedal, A.~Thiel,
  F.~Tufvesson, A.~F. Molisch, and C.~F. Mecklenbr\"auker, ``In-tunnel
  vehicular radio channel characterization,'' in \emph{73rd IEEE Vehicular
  Technology Conference (VTC Spring)}, May 2011.

\bibitem{Thomae2000}
R.~Thom\"a, D.~Hampicke, A.~Richter, G.~Sommerkorn, A.~Schneider, U.~Trautwein,
  and W.~Wirnitzer, ``Identification of time-variant directional mobile radio
  channels,'' \emph{{IEEE} Transactions on Instrumentation and Measurement},
  vol.~49, no.~2, pp. 357--364, April 2000.

\bibitem{RUSK}
{MEDAV}, ``{RUSK} channelsounder,'' http://www.channelsounder.de.

\bibitem{Klemp2010}
A.~Thiel, O.~Klemp, A.~Paiera, L.~Bernado, J.~Karedal, and A.~Kwoczek,
  ``In-situ vehicular antenna integration and design aspects for
  vehicle-to-vehicle communications,'' in \emph{4th European Conference on
  Antennas and Propagation (EuCAP)}, April 2010.

\bibitem{MEDAV}
``{MEDAV GmbH},'' {http://www.medav.de}.

\bibitem{ITS_2008_671_EC}
``Commission decision on the harmonised use of radio spectrum in the 5875-5905
  {MHz} frequency band for safety-related applications of intelligent transport
  systems ({ITS}),'' 2008/671/EC, August 2008.

\bibitem{Paier2010t}
A.~Paier, ``The vehicular radio channel in the 5 {GHz} band,'' Ph.D.
  dissertation, Vienna University of Technology, 2010.

\bibitem{ETSITR102638}
{ETSI TR 102 638}, ``Intelligent transport sytems ({ITS}); vehicular
  communications; basic set of applications; definitions,'' V1.1.1, June 2009.

\bibitem{Bernado2012thesis}
L.~Bernad\'o, ``{Non-Stationarity in Vehicular Wireless Channels},'' Ph.D.
  dissertation, Vienna University of Technology, 2012.

\bibitem{Bernado2012book}
L.~Bernad\'o, N.~Czink, T.~Zemen, A.~Paier, F.~Tufvesson, C.~F.
  Mecklenbr\"auker, and A.~F. Molisch, ``Vehicular channels,'' in \emph{LTE
  Advanced and Beyond Wireless Networks: Channel Modeling and Propagation},
  G.~de~la Roche, A.~Alay{\'o}n, and B.~Allen, Eds.\hskip 1em plus 0.5em minus
  0.4em\relax John Wiley \& Sons Ltd., 2012.

\bibitem{Percival1993}
D.~Percival and A.~Walden, \emph{Spectral analysis for physical applications:
  multitaper and conventional univariate techniques}, ser. Spectral Analysis
  for Physical Applications: Multitaper and Conventional Univariate
  Techniques.\hskip 1em plus 0.5em minus 0.4em\relax Cambridge University
  Press, 1993.

\bibitem{Thomson1982}
D.~Thomson, ``Spectrum estimation and harmonic analysis,'' \emph{Proceedings of
  the {IEEE}}, vol.~70, no.~9, pp. 1055--1096, September 1982.

\bibitem{Slepian1978}
D.~Slepian, ``Prolate spheroidal wave functions, {Fourier} analysis, and
  uncertainty - {V}: The discrete case,'' \emph{The Bell System Technical
  Journal}, vol.~57, no.~5, pp. 1371--1430, May-June 1978.

\bibitem{Bernado2012}
L.~Bernad\'o~and, T.~Zemen, F.~Tufvesson, A.~F. Molisch, and C.~F.
  Mecklenbr\"auker, ``{The (in-)validity of the WSSUS assumption in vehicular
  channels},'' in \emph{23rd IEEE International Symposium on Personal Indoor
  and Mobile Radio Communications (PIMRC)}, September 2012.

\bibitem{Hlawatsch11}
F.~Hlawatsch and G.~Matz, Eds., \emph{Wireless Communications over Rapidly
  Time-Varying Channels}.\hskip 1em plus 0.5em minus 0.4em\relax Academic
  Press, 2011.

\bibitem{Molisch1999}
A.~F. Molisch and M.~Steinbauer, ``Condensed parameters for characterizing
  wideband mobile radio channels,'' \emph{International Journal of Wireless
  Information Networks}, vol.~6, pp. 133--154, 1999, 10.1023/A:1018895720076.

\bibitem{McLachlan2000}
G.~McLachlan and D.~Peel, \emph{Finite Mixture Models}.\hskip 1em plus 0.5em
  minus 0.4em\relax Hoboken, NJ: John Wiley \& Sons, Inc, 2000.

\bibitem{Massey1951}
\BIBentryALTinterwordspacing
F.~J. Massey, ``The {Kolmogorov-Smirnov} test for goodness of fit,''
  \emph{Journal of the American Statistical Association}, vol.~46, no. 253, pp.
  68--78, 1951. [Online]. Available: \url{http://www.jstor.org/stable/2280095}
\BIBentrySTDinterwordspacing

\bibitem{Paier2010}
A.~Paier, L.~Bernad\'o~and, J.~Karedal, O.~Klemp, and A.~Kwoczek, ``Overview of
  vehicle-to-vehicle radio channel measurements for collision avoidance
  applications,'' in \emph{71st IEEE Vehicular Technology Conference (VTC
  Spring)}, May 2010.

\bibitem{Fleury1996}
B.~Fleury, ``{An uncertainty relation for WSS processes and its application to
  WSSUS systems},'' \emph{{IEEE} Transactions on Communications}, vol.~44,
  no.~12, pp. 1632--1634, December 1996.

\end{thebibliography}

\end{document}